\documentclass[10pt]{article}
\usepackage[dvips]{graphicx}
\usepackage{tabularx}
\usepackage{amsmath,amssymb}
\usepackage{bm}
\newcommand{\eq}{\begin{equation}}
\newcommand{\eqend}{\end{equation}}
\newcommand{\ovl}{\overline}
\newcommand{\A}{\alpha}

 \addtocounter{section}{0}
 \makeatletter \@addtoreset{equation}{section}
 \makeatother
 \renewcommand{\theequation}{\arabic{section}.\arabic{equation}}
\title{Minimum Supersymmetric Standard Model on the Noncommutative Geometry}
\author{Satoshi Ishihara, 
  \footnote{E-mail:satoshi@yukawa.kyoto-u.ac.jp}
\and
Hironobu Kataoka,
  \footnote{E-mail:s499756@hyogo-c.ed.jp}
\and
Atsuko Matsukawa,
  \footnote{E-mail:Atsuko Matsukawa@cap.ocn.ne.jp}
\and 
Hikaru Sato,
  \footnote{E-mail:hikaru\underline{ }sato@gakushikai.jp} \\
{\it Department of Physics, Hyogo University of Education} \\
{\it Shimokume, Kato-shi, Hyogo 673-1494, Japan} \\
\\ 
Masafumi Shimojo \footnote{E-mail:shimo0@ei.fukui-nct.ac.jp} \\
{\it Department of Electronics and Information Engineering, }\\
{\it Fukui National College of Technology,} \\
{\it Geshicho, Sabae-Shi, Fukui 916-8507, Japan}
}
\begin{document}
\maketitle
\begin{abstract}
We have obtained the supersymmetric extension of spectral triple which specify a noncommutative geometry(NCG). 
We assume that the functional space $\mathcal{H}$  constitutes of wave functions of matter fields and 
their superpartners included in the minimum supersymmetric 
standard model(MSSM). 
We introduce the internal fluctuations to the Dirac operator on the manifold as well as on the finite space 
by elements of the algebra $\mathcal{A}$ in the triple. So, we obtain not only the vector supermultiplets 
which meditate $SU(3)\otimes SU(2)\otimes U(1)_Y$ gauge degrees of freedom but also Higgs supermultiplets 
which appear in MSSM on the same standpoint. Accoding to the supersymmetric 
version of the spectral action principle, we calculate the square of the fluctuated total Dirac operator 
and verify that the Seeley-DeWitt coeffients give the correct action of MSSM. 
We also verify that the relation between coupling constants of $SU(3)$,$SU(2)$ and $U(1)_Y$ is same as that of 
SU(5) unification theory. 
\end{abstract}
\section{\large INTRODUCTION}
\ \ \ 
Yang-Mills gauge theory which provides the basis of the standard model of high energy physics and 
general relativity theory greatly have succeeded in describing the basic interactions in our Universe.
But in the quantum level, they are difficult to be unified due to the unrenormalizability of gravity. 

The standard model coupled to gravity was derived on the basis of noncommutative geometry(NCG) by 
Connes and his co-workers\cite{rf:Grcoupled,rf:NCGneutrino,rf:JHigh}. 
The framework of NCG is specified by a set called spectral triple $(\mathcal{H},\mathcal{A},\mathcal{D})$\cite{connes0}. 
Here, $\mathcal{A}$ is a noncommutative complex algebra, acting 
on the Hilbert space $\mathcal{H}$, whose elements correspond to spinorial wave functions of physical matter fields, 
while the Dirac operator $\mathcal{D}$ is a self adjoint operator with compact resolvent which play the role of 
metric of the geometry. Higgs, gauge and gravity fields which meditate all basic interactions that we know 
are introduced in the same standpoint, the fluctuations of Dirac operator.   

The internal fluctuation of the Dirac operator is given as follows:
\begin{equation}
\tilde{\mathcal{D}} = \mathcal{D} + A+ JAJ^{-1},\ A=\sum a_i[\mathcal{D},b_i],\ a_i,b_i\in \mathcal{A}.
\end{equation}   
For the Dirac operator on the manifold $\mathcal{D}_M=i\gamma^\mu\nabla_\mu\otimes 1$, the fluctuation 
$A+JAJ^{-1}$ gives the gauge vector fields, while for the Dirac operator in the finite space,$D_F$, it 
gives the Higgs fields\cite{connes7,connes10}. 

The action for the NCG model is obtained by the spectral action principle and expressed by
\begin{equation}
\langle \psi \tilde{D}  \psi \rangle+ {\rm Tr}(f(P)), \label{totalaction0}
\end{equation}
where $\psi$ is a fermionic field which is an element 
in $\mathcal{H}$, $f(x)$ is an auxiliary smooth function on a compact Riemann Manifold 
without boundary of dimension 4. The second term of the action (\ref{totalaction0}) is the bosonic part 
which depends only on the spectrum of the squared Dirac operator $P=\tilde{D}^2$\cite{connes8} and represents 
the non-abelian gauge fields, Higgs fields and gravity. 

In our previous papers\cite{paper0,paper1}, we extended the spectral triple of NCG to a counterpart in the supersymmetric theory 
which may overcome various shortcomings of the standard model\cite{martin} such as hierarchy problem, many 
free parameters to be determined by experiments. We derived the internal fluctuation of $\mathcal{D}_M$, the part of 
Dirac operator acting on manifold, which induced the vector supermultiplets with $U(N)$ internal degrees of freedom. 
Using the modified Dirac operator on the manifold, $\widetilde{\mathcal{D}}_M$, 
we obtained bilinear form similar to the first term in (\ref{totalaction0}) which represented the 
the action of the chiral and antichiral supermultiplets of matter fields and their superpartners.  
Following the supersymmetric version of spectral action principle, 
we calculated the Seeley-DeWitt coefficients due to $\mathcal{D}_M$, and obtained the action for the 
vector supermultiplet which included gauge fields with $U(N)$ internal degrees of freedom. 

In this paper, we assume that the functional space $\mathcal{H}$ is the space of wave functions of 
matter particles and their superpartners which appear in the minmum supersymmetric standard model(MSSM).  
We will introduce the internal fluctuation not only to $\mathcal{D}_M$ but also to the Dirac operator $\mathcal{D}_F$ 
which acts on the finite space.    
The modified total Dirac operator is expressed by
$i\widetilde{\mathcal{D}}_{tot} =i\widetilde{\mathcal{D}}_M \otimes 1_F+ \gamma_M\otimes \widetilde{\mathcal{D}}_F $, 
where $\gamma_M$ is $Z/2$ grading operator on the manifold and 
$\widetilde{\mathcal{D}}_F$ is the modified Dirac operator on the finite space. 
From the fluctuation in $i\widetilde{\mathcal{D}}_M$, we will obtain the vector supermultiplet 
which meditate $U(3)\otimes U(2)\otimes U(1)$ gauge degrees of freedom. We take out $U(1)$'s from $U(3)$ and $U(2)$ 
and combine them with the other $U(1)$ to produce $U(1)$ of weak hypercharge. 
From the fluctuation to $\gamma_M\otimes \widetilde{\mathcal{D}}_F$, we also obtain supermultiplets which transform 
as Higgs fields of MSSM.  

We will calculate the square of the modified total Dirac operator $P=(i\widetilde{\mathcal{D}}_{tot})^2 $, 
and Seeley-DeWitt coefficients of squared total Dirac operator which include the vector supermultiplets as well as 
the Higgs supermultiplets. In section 2, we will review the supersymetric extension of the spectral triple 
which specifies NCG. In section 3, we will choose elements of $\mathcal{A}$ as fluctuations of the total 
Dirac operator to produce vector supermultiplets and 
Higgs supermultiplets which appear in MSSM. In section 4, we 
will calculate the Seeley-DeWitt coefficients due to $P=(i\mathcal{D}_{tot})^2$. 
and verify that we obtain the correct action of MSSM. In the process, we will also obtain the relation between 
coupling constants of $SU(3),SU(2),U(1)_Y$ gauge degrees of freedom which is same as that of 
$SU(5)$ unification theory.     
  
\section{\large THE TRIPLE FOR MSSM}
\ \ \ 
Let us start with reviewing the supersymmetric counterpart $(\mathcal{H},\mathcal{A},\mathcal{D})$\cite{paper1} extended from 
the spectral triple which specifies NCG. 
The functional space $\mathcal{H}$ is the product denoted by
\begin{equation}
\mathcal{H} =\mathcal{H}_M\otimes \mathcal{H}_F. \label{H}
\end{equation}
The functional space $\mathcal{H}_M$ on the the Minkowskian space-time manifold is the direct sum of two subsets, $\mathcal{H}_+$ and 
$\mathcal{H}_-$:
\begin{equation}
\mathcal{H}_M = \mathcal{H}_+ \oplus \mathcal{H}_- \label{HM},
\end{equation}
where $\mathcal{H}_+$ is the space of chiral supermultiplets which constitutes of Weyl spinors which transform as the $(\frac{1}{2},0)$of the Lorentz group $SL(2,C)$ and their superpartners and $\mathcal{H}_-$ is the space of antichiral supermultiplets. The elements 
are expressed by
\begin{align}
\Phi_+ & =((\Psi_+)_i,0^3)^T =(\varphi_+(x),\psi_{+\alpha}(x),F_+(x),0^3)^T \in \mathcal{H}_+, \\
\Phi_- & =(0^3,(\Psi_-)_{\bar{i}})^T = (0^3,\varphi_-^\ast(x),\bar{\psi}_-^{\dot{\alpha}}(x),F_-^\ast(x))^T \in\mathcal{H}_-.
\end{align}
$Z_2$ grading $\gamma_M$ on the space $\mathcal{H}_M$ is given by 
\begin{equation}
\gamma_M(\Phi_+)=-i,\ \ \gamma_M(\Phi_-) = i.
\end{equation}

A supersymmetric invariant product of wave functions $\Psi$,$\Psi^\prime$ is  defined by
\begin{equation}
(\Psi^\prime,\Psi) =\int \Psi^{\prime\dagger}\Gamma_0 {\rm \Psi} d^4x, \label{suinvproduct}
\end{equation} 
where $\Psi$ is expressed on the base by
\begin{equation}
\Psi=\begin{pmatrix}
\Psi_+\\
\Psi_-
\end{pmatrix}=\Phi_+ + \Phi_-,
\end{equation}
${\rm \Gamma}_0$ is given by
\begin{equation}
\Gamma_0 = \begin{pmatrix}
0 & \mathit{\Gamma}_0\\
\mathit{\Gamma}_0 & 0
\end{pmatrix},\ \ 
\mathit{\Gamma}_0 =\begin{pmatrix}
0 & 0 & 1\\
0 & -1 & 0\\
1 & 0 & 0
\end{pmatrix}. 
\end{equation}

The algebra $\mathcal{A}_M=\mathcal{A}_+ \oplus \mathcal{A}_-$ which acts on $\mathcal{H}_M$ constitutes of elements 
expressed by
\begin{align}
u_a = \left(\begin{array}{@{\,} c|c @{\,}} 
(u_a)_{ij} & 0 \\ \hline 
0 & 0
\end{array}
\right) \in \mathcal{A}_+,\ \ 
(u_a)_{ij} & = \frac{1}{m_0}
\begin{pmatrix}
\varphi_a & 0 & 0\\
\psi_{a\alpha} & \varphi_a & 0 \\
F_a & -\psi_a^\alpha & \varphi_a
\end{pmatrix} , 
\label{ua}
\\
\ovl{u}_a = \left(\begin{array}{@{\,} c|c @{\,}}
0 & 0 \\ \hline
0 & (\ovl{u}_a)_{\bar{i}\bar{j}}
\end{array}
\right) \in \mathcal{A}_-,\ \ 
(\bar{u}_a)_{\bar{i}\bar{j}} & = \frac{1}{m_0}
\begin{pmatrix}
\varphi_a^\ast & 0 & 0\\
\bar{\psi}_a^{\dot{\alpha}} & \varphi_a^\ast & 0 \\
 F_a^\ast & -\bar{\psi}_{a\dot{\alpha}} & \varphi_a^\ast
\end{pmatrix} \label{barua},
\end{align}
where $\{\varphi_a(\varphi_a^\ast), \psi_{a\alpha}(\bar{\psi}_a^{\dot{\alpha}}),F_a(F_a^\ast)\}$
is a chiral(antichiral) supermultiplet. We note that $\mathcal{A}_\pm$ includes the constant functions expressed by
\begin{equation}
1_+=\left(\begin{array}{c|c}
\delta_{ij} & 0\\
\hline
0 & 0
\end{array}\right)\in \mathcal{A}_+,\ 
1_-=\left(\begin{array}{c|c}
0 & 0\\
\hline
0 & \delta_{\bar{i}\bar{j}}
\end{array}\right)\in \mathcal{A}_-
\label{constantu}
\end{equation}

The Dirac operator which acts on the $\mathcal{H}_M$ is given by
\begin{equation}
\mathcal{D}_M = -i\begin{pmatrix}
0 & \bar{\mathcal{D}}_{i\bar{j}} \\
\mathcal{D}_{\bar{i}j} & 0
\end{pmatrix}, 
\label{DM0}
\end{equation}
where
\begin{equation}
\mathcal{D}_{\bar{i}j}=\begin{pmatrix}
0 & 0 & 1\\
0 & i\bar{\sigma}^\mu\partial_\mu & 0\\
\Box & 0 & 0
\end{pmatrix},\ 
\bar{\mathcal{D}}_{i\bar{j}}= \begin{pmatrix}
0 & 0 & 1\\
0 & i\sigma^\mu\partial_\mu & 0\\
\Box & 0 & 0
\end{pmatrix}.
\label{DM}
\end{equation}

The algebra $\mathcal{A}_F$ on the finite space $\mathcal{H}_F$ is given by
\begin{equation}
\mathcal{A}_F = \mathbb{C}\oplus \mathbb{H}\oplus M_3(\mathbb{C}),
\end{equation}
where $\mathbb{H}$ is the space of quaternions and $M_3(\mathbb{C})$ is the space of $3\times 3$ complex matrices. 
The representation space of $\mathcal{A}_F$, $\mathcal{H}_F$ is the space of labels which denote quantum 
numbers of quarks and leptons as follows:
\begin{equation}
Q_{AI} = \begin{pmatrix}
(u_L)_A \\
(d_L)_A
\end{pmatrix},\  (u_R)_A,\ (d_R)_A,\ 
l_I =\begin{pmatrix}
\nu_L\\
e_L
\end{pmatrix}_I,\ e_R, \label{particles}
\end{equation}
and labels of their antiparticles are also elements of $\mathcal{H}_F$ which are expressed by
\begin{align}
&Q^c = \begin{pmatrix}
(u^c)_R \\
(d^c)_R
\end{pmatrix} = \begin{pmatrix}
(u_L)^* \\
(d_L)^*
\end{pmatrix} = Q^*, \mbox{ }  
\nonumber \\
&(u^c)_L = (u_R)^*, \mbox{ } (d^c)_L = (d_R)^* ,
\nonumber \\
&\ell^c = \begin{pmatrix}
(\nu^c)_R \\
(e^c)_R
\end{pmatrix} =
\begin{pmatrix}
(\nu_L)^* \\
(e_L)^*
\end{pmatrix} = \ell^*, \mbox{ } (e^c)_L = (e_R)^*,
\label{eq1.6}
\end{align}
where $A$ denotes indices on which $3\times 3$ complex matrices act and $I$ denotes those on which quaternions act. 
$L$ and $R$ denote that the eigenvalue of $Z/2$ grading $\gamma_F$ is $-1$ and $1$, respectively.

We define the action of $a \in \mathcal{A}_F$ for the quark sector as follows:
\begin{align}
a \begin{pmatrix}
u_R \\
d_R
\end{pmatrix} &=
\begin{pmatrix}
\lambda^{m_1} & 0 \\
0 & {{\lambda}^{m_2}}
\end{pmatrix} 
\begin{pmatrix}
u_R \\
d_R
\end{pmatrix}, 
\mbox{ } a \begin{pmatrix}
u_L \\
d_L
\end{pmatrix}_I =
q_I^{\ I^\prime} 
\begin{pmatrix}
u_L \\
d_L
\end{pmatrix}_{I^\prime}
\label{aquark}
\end{align}
\begin{equation}
a \begin{pmatrix}
Q^{cA} \\
u^{cA} \\
d^{cA}
\end{pmatrix} = 
\begin{pmatrix}
{(m^*)^A}_B & 0 & 0 \\
0 & {(m^*)^A}_B & 0 \\
0 & 0 & {(m^*)^A}_B 
\end{pmatrix}
\begin{pmatrix}
Q^{cB} \\
u^{cB} \\
d^{cB}
\end{pmatrix},
\label{aantiquark}
\end{equation}
where $\lambda \in \mathbb{C}$, $q\in\mathbb{H}$ and $m\in M_3(\mathbb{C})$. When we denote the base of quark sector as 
\begin{equation}
\begin{pmatrix}
\Psi_q \\
\Psi_q^c
\end{pmatrix},\ \  
\Psi_q=(Q,u_R,d_R)^T,\ \Psi_q^c = (Q^c,(u_R)^c,(d_R)^c)^T,
\end{equation}
$a$ is represented by a matrix given by
\begin{equation} 
a= \left(
\begin{array}{@{\,}ccc|ccc@{\,}}
q \otimes {\bf 1}_3 &  &  & & &  \\
 & \lambda^{m_1} \otimes {\bf 1}_3 &  & & {\bf 0} & \\
 &  & {\bf 1}_3\otimes{{\lambda}^{m_2}}   & & & \\ \hline
 & & & m^*\otimes {\bf 1}_2 &  &  \\
 & {\bf 0} & &  & m^* &  \\
 & & &  &  & m^* 
\end{array}
\right), \label{actionforquark}
\end{equation}
where $\bf{1}_3$ denotes the unit matrix of $M_3(\mathbb{C})$, $\bf{1}_2$ is that of $\mathbb{H}$ 
and $m_i$ are integers. \\

For the lepton sector, we take the base expressed by
\begin{equation}
\begin{pmatrix}
\Psi_l \\
\Psi_l^c
\end{pmatrix},\ \ 
\Psi_l = (
l,e_R)^T, \mbox{ } \Psi^c_\ell = (
\ell^c, (e^c)_L)^T
\end{equation}
and define $a$ as follows:
\begin{equation}
a = \left(
\begin{array}{@{\,}c|c@{\,}}
 \begin{array}{cc}
 q & \\
   & {{\lambda}^{m_3}}
 \end{array} & 
 {\bf 0} \\
\hline
{\bf 0} & 
\begin{array}{cc}
1 & \\
  & 1
\end{array} 
\end{array}
\right)
\label{actionforlepton}
\end{equation}

For an element $a\in \mathcal{A}_F$, we define the real structure as a antilinear isometry  of 
$\mathcal{H}_F$ as follows:
\begin{equation}
a \, \longmapsto J_F \, a^* \, J_F^{-1}.
\label{eq1.12}
\end{equation}
$J_F$ satisfies the following conditions:
\begin{align}
& [a, J_F \,b^* \,J_F^{-1} ] = 0,  \mbox{   }  \forall a, b \in \mathcal{A}_F
\label{eq1.13} \\
& \left[[D_F, a], J_F \,b^* \,J_F^{-1} \right] = 0.
\end{align}
$J_F$ define the right action of $\mathcal{A}_F$ in $\mathcal{H}_F$ as follows:
\eq 
\Psi a^T = J_F a^* J_F^{-1} \Psi, \mbox{ } \Psi \in \mathcal{H}_F.
\label{rightaction}
\eqend
From (\ref{rightaction}), we see that 
the left action of $a$ for the antiparticle determine the right action for the particle and 
can be moved into the left action through $J_F$. For an example, let us consider $Q_{AI}$ in (\ref{particles}). 
From (\ref{actionforquark}), the left and right actions of $\mathcal{A}_F$ on $Q$ are summarized by
\begin{align}
Q^\prime_{AI} & = q_I^{\ J}Q_{BJ}\left((J_F m^\ast J_F^{-1})^T\right)^B_{\ \ A}= q_I^{\ J}m_A^{\ B}Q_{BJ}.
\end{align} 

The Dirac operator $\mathcal{D}_F$ on $\mathcal{H}_F$ is given for each label of $s=u,d,e$ by 
\begin{equation}
\mathcal{D}_F =
\begin{pmatrix}
0 & m_s^\dagger\\
m_s & 0
\end{pmatrix}, \label{DF}
\end{equation}
where $m_s$ is the mass matrix with respect to the family index.

In order to evade fermion doubling, we impose the following condition
\begin{equation}
\gamma_M\gamma_F= i,
\end{equation}
and extract the physical wave functions as follows: \\
for quarks
\begin{equation}
(\Phi_+ \oplus \Phi_-) \otimes \begin{pmatrix}
u_L \\
d_L \\
u_R (d_R)
\end{pmatrix}
 \xrightarrow{\gamma_M \gamma_F = i}
 \begin{pmatrix}
 \begin{pmatrix}
 \Psi_+ \\ 
  0
 \end{pmatrix} 
 \otimes \begin{pmatrix}
 u_L \\
 d_L
 \end{pmatrix} \\
 \begin{pmatrix}
 0\\ 
 \Psi_- 
 \end{pmatrix} \otimes u_R (d_R) \\
 \end{pmatrix},
\label{physicalquark} 
\end{equation}
and for leptons
\begin{equation}
(
\Phi_+ \oplus \Phi_-
)
\otimes \begin{pmatrix}
\nu_L \\
e_L \\
e_R 
\end{pmatrix}
 \xrightarrow{\gamma_M \gamma_F = i}
 \begin{pmatrix}
 \begin{pmatrix}
 \Psi_+ \\
 0
 \end{pmatrix}
 \otimes \begin{pmatrix}
 \nu_L \\
 e_L
 \end{pmatrix} \\
 \begin{pmatrix}
 0\\
 \Psi_- 
 \end{pmatrix}
 \otimes e_R \\
 \end{pmatrix},\label{physicallepton}
\end{equation}

From (\ref{ua}),(\ref{barua}),(\ref{actionforquark}),(\ref{actionforlepton}), 
we summarize in Table 1 the action of the elements of the 
algebra $\mathcal{A}_M\otimes \mathcal{A}_F$ on the physical states. 
\begin{table}
\begin{center}
\begin{tabular}[h]{lll}
{\rm particle} $s$ & {\rm element\ $u_{a0}^{(s)}$} & {\rm $u_{a1}^{(s)}=Ju_a^{(s^c)}J^{-1} $} \\
{\rm quark\ sector}\\
$Q_\alpha(x)$ & $u_a^{[2]}=(q(x))_I^{\ J}\in \mathbb{H}$ & $u_a^{[3]}=(m(x))_A^{\ B}\in M_3(\mathbb{C}) $ \\
$u^{\dot{\alpha}}_R(x)$ & $u_{a}^{[1_1]}=\lambda^{m_1}(x)\in \mathbb{C}$  &  $u_a^{[3]}$ \\
$ d^{\dot{\alpha}}_R(x)$ & $u_{a}^{[1_2]}=\lambda^{m_2}(x)\in \mathbb{C}$ 
& $u_a^{[3]}$\\
{\rm lepton\ sector} \\
$l_\alpha(x)$ & $u_a^{[2]}$
& $1$ 
\\
$e_R^{\dot{\alpha}}(x)$ & $u_{a}^{[1_3]}=\lambda^{m_3}(x)\in \mathbb{C}$ 
& $1$ 
\end{tabular}
\caption{The list of matter fields and the action of the algebra for them:
The third column denotes the actions of elements of the algebra moved from the antiparticle part through the 
real structure $J$.}
\end{center}
\end{table}
The matrix form of these elements on the the basis (\ref{physicalquark}) 
is expressed by
\begin{align}
U_a & = \left(\begin{array}{@{\,} c|c @{\,}} 
u_a^{(Q)} & 0 \\ \hline 
0 & u_a^{(u_R(d_R))}
\end{array}
\right)
\in \mathcal{A}_+\otimes \mathcal{A}_F, \label{Ua}\\
\ovl{U}_a & = \left(\begin{array}{@{\,} c|c @{\,}} 
\bar{u}_a^{(Q)} & 0 \\ \hline 
0 & \bar{u}_a^{(u_R(d_R))}
\end{array}
\right)
\in \mathcal{A}_-\otimes \mathcal{A}_F, \label{barUa}
\end{align}
where $u_a^{(s)},s=Q,u_R(d_R)$ are elements of $\mathcal{A}_+$ in the form of (\ref{ua}) 
and they are linear combinations of $u_{a0}^{(s)}$ and $u_{a1}^{(s)}$ in the Table 1, while 
$\bar{u}_a^{(s)}$ are elements of $\mathcal{A}_-$ in the form of (\ref{barua}) and they are 
related to  $u_a^{(s)}$ as follows:
\begin{equation}
(\mathit{\Gamma}_0 u_a)^\dagger = \mathit{\Gamma}_0 \bar{u}_a
\end{equation} 
  
Constant elements of $\mathcal{A}_\pm\otimes \mathcal{A}_F$ are also given by
\begin{equation}
\mathbf{1}_+ = 
\left(
\begin{array}{c|c}
1_+^{(Q)} & 0\\
\hline
0 & 1_+^{(u_R(d_R))}
\end{array}
\right),\ 
\mathbf{1}_- = 
\left(
\begin{array}{c|c}
1_-^{(Q)} & 0\\
\hline
0 & 1_-^{(u_R(d_R))}
\end{array}
\right),\ 
\label{constantU}
\end{equation}
where $1_\pm^{(s)}$ denotes the matrix form of (\ref{constantu}). 

If we replace $Q\rightarrow l$ and $u_R(d_R)\rightarrow e_R$ in (\ref{Ua}),(\ref{barUa}),(\ref{constantU}), 
we obtain formulas for the lepton sectors. Hereafter, we discuss the quark sector principally and obtain formulas for  
the lepton sector through this replacement.

On the basis, the total Dirac operators for the quark sector 
is given as follows:
\begin{align}
i\mathcal{D}_{tot} & = i{\mathcal{D}}_M \otimes \mathbf{1}_F + \gamma_M \otimes \mathcal{D}_F
\nonumber \\
& = \left(
\begin{array}{@{\,}cc|cc@{\,}}
 0  & \ovl{\mathcal{D}} &  
0 & 0  \\
 \mathcal{D} & 0 & 
 0 & 
 iM^{(u(d))\dagger} \\
 \hline
-iM^{(u(d))} 
& 0  
& 0  & \ovl{\mathcal{D}} \\
0 & 
0 & 
\mathcal{D} & 0
\end{array} \right), \label{iDtot}
\end{align}
where $M^{(u(d))}$ represents the doublet term which includes mass of quarks expressed by
\begin{equation}
M^{(u)} =(m_u,0)=y_u(\mu_0,0),\ \ M^{(d)}=(0,m_d)=y_d(0,\mu_0), \label{Mud} 
\end{equation}
where $y_u$ and $y_d$ are Yukawa matrices with regard to generation. 

For the antiparticle sector, the total Dirac operator is given by 
\begin{equation}
J i\mathcal{D}_{tot}J^{-1},
\end{equation}
where $J$ is the real structure of $\mathcal{H}=\mathcal{H}_M\otimes \mathcal{H}_F$. 

We note that under the definition of supersymmetric invariant product (\ref{suinvproduct}), $i\mathcal{D}_{tot}$ has 
the following hermiticity: 
\begin{equation}
({\rm \Gamma}_0i\mathcal{D}_{tot})^\dagger ={\rm \Gamma}_0i\mathcal{D}_{tot}. \label{hermiteDtot}
\end{equation}
\section{\normalsize INTERNAL FLUCTUATION, VECTOR AND  HIGGS SUPERMULTIPLET}
\ \ \ 
In the theory of noncommutative geometry without supersymmetry, 
gauge fields and Higgs fields are derived through internal fluctuation of Dirac operator in the form expressed by
\begin{equation}
\sum_j a_j[D,b_j] + \sum_j Ja_j[D,b_j]J^{-1},\ \ a_j,b_j\in \mathcal{A}. \label{JAJ}.
\end{equation} 
The second term of (\ref{JAJ}) is the fluctuation given by the elements of algebra which 
act on the antiparticle part and transferred to the particle part through the real structure $J$. 
And gauge fields are derived from the fluctuations for the Dirac operator which acts on the 
Riemann manifold , while the Higgs fields are given by those for the Dirac operator on the finite space.

We will consider to extend these constructions to supersymmetric version. The major differences from 
the non-supersymmetric case is due to the fact that 
the algebra which is the source of the fluctuation is the direct sum of 
the spaces $\mathcal{A}_\pm\otimes \mathcal{A}_F$.

Using (\ref{Ua}),(\ref{barUa}),(\ref{constantU}),(\ref{iDtot}), we obtain the following expressions:
\begin{align}
\ovl{U}_a[i\mathcal{D}_{tot},U_a] & = 
\left(\begin{array}{cc|cc}
0 & 0 & 0 & 0 \\
\bar{u}_{a\bar{i}\bar{j}}^{(Q)} \mathcal{D}_{\bar{j}k} u_{akl}^{(Q)} & 0 & 0 & 0\\
\hline
0 & 0 & 0 & 0\\
0 & 0 & \bar{u}_{a\bar{i}\bar{j}}^{(s)} \mathcal{D}_{\bar{j}k} u_{akl}^{(s)} & 0
\end{array}\right) \label{barUDU}
\end{align}
\begin{align}
U_a[i\mathcal{D}_{tot},\ovl{U}_a] & =
\left(\begin{array}{cc|cc}
0 & u_{aij}^{(Q)} \ovl{\mathcal{D}}_{j\bar{k}} \bar{u}_{a\bar{k}l}^{(Q)} & 0 & 0\\
0 & 0 & 0 & 0\\
\hline
0 & 0 & 0 & u_{aij}^{(s)} \ovl{\mathcal{D}}_{j\bar{k}} \bar{u}_{a\bar{k}\bar{l}}^{(s)} \\
0 & 0 & 0 & 0
\end{array}\right), \label{UDbarU}
\end{align}
\begin{align}
U_a[i\mathcal{D}_{tot},U_a]-U_aU_a[i\mathcal{D}_{tot},\mathbf{1}_+] & = 
\left(
\begin{array}{cc|cc}
0 & 0 & 0 & 0\\
0 & 0 & 0 & 0\\
\hline
-i(u_a^{(s)}M^{(u(d))}u_a^{(Q)}-u_a^{(s)}u_a^{(s)}M^{u(d)})_{ij}& 0 & 0 & 0 \\
0 & 0 & 0 & 0
\end{array}
\right) \label{fDF}
\\
[i\mathcal{D}_{tot},\ovl{U}_a]\ovl{U}_a-[i\mathcal{D}_{tot},\mathbf{1}_-]\ovl{U}_a\ovl{U}_a & =
\left(
\begin{array}{cc|cc}
0 & 0 & 0 & 0\\
0 & 0 & 0 & i(M^{(u(d))\dagger} \bar{u}_a^{(s)}\bar{u}_a^{(s)}-\bar{u}_a^{(Q)}M^{(u(d))\dagger} \bar{u}_a^{(s)})_{\bar{i}\bar{j}} \\
\hline
0 & 0 & 0 & 0 \\
0 & 0 & 0 & 0
\end{array}
\right), \label{fDFMd}
\end{align}
where $s=u_R(d_R)$. 
Eq. (\ref{barUDU}) and (\ref{UDbarU}) give fluctuations only to $i\mathcal{D}_M\otimes \mathbf{1}_F$, 
so we will see later that this type of 
fluctuation  produce vector superfields, while Eq.(\ref{fDF}) and (\ref{fDFMd}) give contribution only to 
$\gamma_M\otimes \mathcal{D}_F$ so that this type of fluctuation will give Higgs superfields.
%
%

\subsection{\normalsize VECTOR SUPERMULTIPLET}
\ \ \ 
We introduce a set of elements of $\mathcal{A}$ in (\ref{Ua}),(\ref{barUa}) expressed by
\begin{align}
\Pi_+ &= \{ u_a^{[r]} ; a = 1, 2, \cdots, n_r \} \subset \mathcal{A}_+ \otimes \mathcal{A}_F, 
\label{Pi+}
\\
{\Pi}_- &=  \{ \ovl{u}_a^{[r]} ; a = 1, 2, \cdots, n_r \} \subset \mathcal{A}_- \otimes \mathcal{A}_F,
\label{Pi-}
\end{align}
where $r=1,2,3$  and 
$u_a^{[r]},\bar{u}_a^{[r]} $ are in the matrix form of (\ref{ua}),(\ref{barua}) and given in the Table 1. 
The elements of 
$\Pi_+$ and $\Pi_-$ are chosen such that the products of $u_a^\prime s$ and $\bar{u}_a^\prime s$ 
do not belong to $\Pi_+,\Pi_-$ any more. 

We can define elements $(A_\mu^{[r]}, \lambda_\alpha^{[r]},D^{[r]})$ of a vector supermultiplet  as follows:
\begin{align}
& m_0^2 \, A_\mu^{[r]} = i \sum_a c_a^{[r]} \left[(\varphi_a^{[r]*} \partial_\mu \varphi_a^{[r]} - \partial_\mu \varphi_a^{[r]*} \varphi_a^{[r]} ) - i \ovl{\psi}_{a\dot{\A}}^{[r]}\ovl{\sigma}_\mu^{\dot{\A}\A} \psi_{a\A}^{[r]} \right],
\label{Amu} 
\\
&m_0^2 \,  \lambda_\A^{[r]} = -\sqrt{2} i \sum_a c_a^{[r]} \left( F_a^{[r]*} \psi_{ a\A}^{[r]} - i \sigma^\mu_{\A\dot{\A}} \ovl{\psi}_a^{[r]\dot{\A}} \partial_\mu \varphi_a^{[r]} \right),
\label{eq2.14d}
\\
&m_0^2 \, D^{[r]} =  -\sum_a c_a^{[r]} \left[-2(\partial^\mu \varphi_a^{[r]*} \partial_\mu \varphi_a^{[r]}) + i  \left\{ \partial_\mu \ovl{\psi}_{a\dot{\A}}^{[r]} \ovl{\sigma}^{\mu \dot{\A}\A} \psi_{a\A}^{[r]} -\ovl{\psi}_{a\dot{\A}}^{[r]} \ovl{\sigma}^{\mu \dot{\A}\A} \partial_\mu \psi_{a\A}^{[r]} \right\} + 2  F_a^{[r]*} F_a^{[r]} \right],
\label{D}
\end{align}
where $c^{[r]}$ are the real coefficients 
and elements $(C^{[r]},\chi_\alpha^{[r]},M^{[r]},N^{[r]})$ can be defined as follows:
\begin{align}
&m_0^2 \, C^{[r]} = -\sum_a c_a^{[r]} \varphi_a^{[r]*} \varphi_a^{[r]},
\label{C} \\ 
& m_0^2 \, \chi_\A^{[r]} = i \sqrt{2} \sum_a c_a^{[r]} \varphi_a^{[r]*} \psi_{a\A}^{[r]},
\\
&  m_0^2 \ovl{\chi}^{[r]\dot{\A}} = -i \sqrt{2} \sum_a c_a^{[r]} \ovl{\psi}_a^{[r]\dot{\A}} \varphi_a^{[r]}, 
\label{eq2.17d} \\
& m_0^2 \, (M + i N)^{[r]} = 2i \sum_a c_a^{[r]} \varphi_a^{[r]*} F_a^{[r]},
\\
&  m_0^2 (M - i N)^{[r]} = -2 i \sum_a c_a^{[r]} F_a^{[r]*} \varphi_a^{[r]}.
\label{M-iN}
\end{align} 

The condition for the fields in (\ref{C})$\sim$ (\ref{M-iN}) to vanish is Wess-Zumino condition and given by
\eq
\begin{cases}
\sum_a c_a^{[r]} \varphi^{[r]*}_a \varphi_a^{[r]} = 0, \\
\sum_a c_a^{[r]} \varphi^{[r]*}_a \psi_a^{[r]\A} = 0, \\
\sum_a c_a^{[r]} \varphi^{[r]*}_a F_a^{[r]} = 0.
\end{cases} 
\label{WessZumino}
\eqend
From (\ref{ua}),(\ref{barua}),(\ref{DM}) and (\ref{Amu})$\sim$(\ref{D}) with Wess Zumino condition (\ref{WessZumino}), we obtain the following 
expressions:
\begin{align}
\sum_a c_{a_r}^{[r]} \bar{u}_{a\bar{i} \bar{k}}^{[r]} \, \mathcal{D}_{\bar{k} \ell} \, u_{a\ell j}^{[r]}  
&= \frac{1}{2} \begin{pmatrix}
0 & 0 & 0 \\
-{i}{\sqrt{2}} \ovl{\lambda}^{[r]\dot{\A}} &  \ovl{\sigma}^{\mu\dot{\A}\A} A_\mu^{[r]} & 0 \\
- D^{[r]} - i\partial^\mu A_\mu^{[r]}  - 2i A_\mu^{[r]} \partial^\mu & -{i}{\sqrt{2}} \lambda^{[r]\A} & 0
\end{pmatrix},
\label{baruDu}
\end{align}
\eq
\sum_a c_{a_r}^{[r]} (u_a)_{i k}^{[r]} \, \ovl{\mathcal{D}}_{k \bar{\ell}} \, (\bar{u}_a)_{\bar{\ell} \bar{j}}^{[r]} =  -\frac{1}{2} \begin{pmatrix}
0 & 0 & 0 \\
-{i}{\sqrt{2}} {\lambda}_{{\A}}^{[r]} &  {\sigma}^{\mu}_{\A\dot{\A}} A_\mu^{[r]} & 0 \\
 D^{[r]} - i\partial^\mu A_\mu^{[r]}  - 2i A_\mu^{[r]} \partial^\mu & -{i}{\sqrt{2}} \ovl{\lambda}_{\dot{\A}}^{[r]} & 0
\end{pmatrix}.
\label{eq2.12c}
\eqend
\begin{align}
\sum_{a_r,a_{r^\prime}} c_{a_r}^{[r]} c_{a_{r^\prime}}^{[r^\prime]} (\ovl{u}_{a_ra_{r^\prime}})^{[rr^\prime]}_{\bar{i}\bar{k}} 
 \mathcal{D}_{\bar{k}l} (u_{a_ra_{r^\prime}})_{\ell j}^{[rr^\prime]} &= -\frac{1}{2} A_\mu^{[r]} A^{[r^\prime]\mu}\delta_{\bar{i}\bar{3}}\delta_{j1}
\\
\sum_{a_r,a_{r^\prime}} c_{a_r}^{[r]}c_{a_{r^\prime}}^{[r^\prime]} ({u}_{a_ra_{r^\prime}})_{{i}{k}}^{[rr^\prime]} \ovl{\mathcal{D}}_{{k} \bar{l}} (\ovl{u}_{a_ra_{r^\prime}})_{ \ovl{\ell} \bar{j}}^{[rr^\prime]} &= -\frac{1}{2} A_\mu^{[r]} A^{[r^\prime]\mu}\delta_{i3}\delta_{\bar{j}\bar{1}},
\label{ubarDbaru}
\end{align} 
where $u_{a_ra_{r^\prime}}^{[rr^\prime]} = u_{a_r}^{[r]}u_{a_{r^\prime}}^{[rr^\prime]}$ and 
$\bar{u}_{a_ra_{r^\prime}}^{[rr^\prime]} = \bar{u}_a^{[r]}\bar{u}_a^{[r^\prime]}$. 

In the Table 1, we see that the algebra which act on $Q_\A$ field is the space of functions in $\mathbb{H}\oplus M_3(\mathbb{C})$.
Let us assume that we choose elements of $\mathcal{A}$ in (\ref{Ua}) in the quark sector 
as follows:
\begin{equation}
U_{a_r}^{[r]} = \left(\begin{array}{c|c}
u_{a_r}^{[r]} & 0\\
\hline
0 & 0
\end{array}\right),\ \ 
U_{a_ra_{r^\prime}}^{[r,r^\prime]}=\left(\begin{array}{c|c}
u_{a_r}^{[r]}u_{a_{r^\prime}}^{[r^\prime]} & 0\\
\hline
0 & 0
\end{array}\right), \ r,r^\prime =2,3. \label{UQ}
\end{equation}
The fluctuation for $\mathcal{D}$, 
$\ovl{\mathcal{D}}$ which produces vector supermultiplet  is 
given by 
\begin{align}
V & = 2\sum_{r=2}^3 \sum_{a_r} c_{a_r}^{[r]} \bar{U}_{a_r}^{[r]}[i\mathcal{D}_{tot}, U_{a_r}^{[r]}] 
+ 2\sum_{r=2}^3\sum_{r^\prime =2}^3 \sum_{a_r,a_{r^\prime}} c_{a_r}^{[r]} c_{a_{r^\prime}}^{[r^\prime]} 
\ovl{U}_{a_ra_{r^\prime}}^{[rr^\prime]} 
[ i\mathcal{D}_{tot}, U_{a_ra_{r^\prime}}^{[rr^\prime]} ] 
\nonumber \\
 & \mbox{ } - 2\sum_{r=2}^3\sum_{a_r} c_{a_r}^{[r]} U_{a_r}^{[r]} [i\mathcal{D}_{tot}, \bar{U}_{a_r}^{[r]}] 
+ 2\sum_{r=2}^3\sum_{r^\prime =2}^3 \sum_{a_r,a_{r^\prime}} c_{a_r}^{[r]} c_{a_{r^\prime}}^{[r^\prime]} 
 {U}_{a_ra_{r^\prime}}^{[rr^\prime]} [i \mathcal{D}_{tot}, \ovl{U}_{a_ra_{r^\prime}}^{[rr^\prime]} ]
\nonumber \\
& = 
\left(\begin{array}{c|c}
 \begin{array}{@{\,} c|c @{\,}}
 0 & (\ovl{V}_{WZ}^{(Q)})_{{i}\bar{j}}\\ \hline 
 ( V_{WZ}^{(Q)})_{\bar{i}{j}} & 0
  \end{array} 
  & \bf{0} \\
   \hline
  \bf{0} & \bf{0}
\end{array}\right). \label{VforQ}
\end{align}
Hereafter, for simplicity, we abbreviate the upper index $[r]$of $c_a^{[r]}$ in (\ref{VforQ}) as far as there is not confusion. 
Using (\ref{baruDu})$\sim$(\ref{ubarDbaru}), the modified  $\mathcal{D}$ and $\ovl{\mathcal{D}}$ 
by these fluctuations  
amount to
\begin{align}
\tilde{\mathcal{D}}_Q & =\mathcal{D}+V_{WZ}^{(Q)} 
=\begin{pmatrix}
0 & 0 & 1\\
-i\sqrt{2}\ovl{\lambda}^{(Q)\dot{\A}} & i\bar{\sigma}\mathcal{D}_\mu^{(Q)} & 0\\
\mathcal{D}_\mu^{(Q)}\mathcal{D}^{(Q)\mu} -D^{(Q)} & -i\sqrt{2}\lambda^{(Q)\A} & 0
\end{pmatrix}, \label{tDQ}
\end{align}
\begin{align}
\tilde{\ovl{\mathcal{D}}}_Q & =\ovl{\mathcal{D}}+\ovl{V}_{WZ}^{(Q)} 
=\begin{pmatrix}
0 & 0 & 1\\
-i\sqrt{2}\lambda_\A^{(Q)} & i\sigma\mathcal{D}_\mu^{(Q)} & 0\\
\mathcal{D}_\mu^{(Q)}\mathcal{D}^{(Q)\mu} +D^{(Q)} & -i\sqrt{2}\lambda_{\dot{\A}}^{(Q)} & 0
\end{pmatrix}, \label{tbarDQ}
\end{align}
where $\mathcal{D}_\mu^{(Q)}$ is the covariant derivatives 
\begin{equation}
\mathcal{D}_\mu^{(Q)}=\partial_\mu -iA_\mu^{(Q)},\ \ A_\mu^{(Q)} =A_\mu^{[2]}+A_\mu^{[3]},
\end{equation}
and 
\begin{equation}
\lambda_\A^{(Q)}=\lambda_\A^{[3]}+\lambda_\A^{[2]},\ D^{(Q)}=D^{[3]}+D^{[2]}.
\end{equation}
The algebra which acts on $u_R(d_R)$ field is the space of functions in $\mathbb{C}\oplus M_3(\mathbb{C})$. 
If we choose the following elements of $\mathcal{A} $ in (\ref{UQ}):
\begin{equation}
U_{a_r}^{[r]} = \left(\begin{array}{c|c}
0 & 0\\
\hline
0 & u_{a_r}^{[r]}
\end{array}\right),\ \ 
U_{a_ra_{r^\prime}}^{[r,r^\prime]}=\left(\begin{array}{c|c}
0 & 0\\
\hline
0 & u_{a_r}^{[r]}u_{a_{r^\prime}}^{[r^\prime]}
\end{array}\right), \ r,r^\prime =1_1(1_2),3, \label{UuRdR}
\end{equation}
instead of (\ref{tDQ}),(\ref{tbarDQ}), we obtain the following expressions:
\begin{equation}
\tilde{\mathcal{D}}_s = 
\begin{pmatrix}
0 & 0 & 1\\
-i\sqrt{2}\ovl{\lambda}^{(s)\dot{\A}} & i\bar{\sigma}\mathcal{D}_\mu^{(s)} & 0\\
\mathcal{D}_\mu^{(s)}\mathcal{D}^{(s)\mu} -D^{(s)} & -i\sqrt{2}\lambda^{(s)\A} & 0
\end{pmatrix},\ \ 
\tilde{\ovl{\mathcal{D}}}_s = 
\begin{pmatrix}
0 & 0 & 1\\
-i\sqrt{2}\lambda_\A^{(s)} & i\sigma\mathcal{D}_\mu^{(s)} & 0\\
\mathcal{D}_\mu^{(s)}\mathcal{D}^{(s)\mu} +D^{(s)} & -i\sqrt{2}\lambda_{\dot{\A}}^{(s)} & 0
\end{pmatrix}, \label{DuRdR}
\end{equation}
where $s=u_R(d_R)$ and 
\begin{equation}
A_\mu^{(s)} =A_\mu^{[3]}+A_\mu^{[1_1(1_2)]},\ 
\lambda_\A^{(s)}=\lambda_\A^{[3]}+\lambda_\A^{[1_1(1_2)]},\ D^{(s)}=D^{[3]}+D^{[1_1(1_2)]}.
\end{equation}
We show the detail of the above operations in Appendix A.
We also show in Appendix A that $(A_\mu^{[r]},\lambda_\alpha^{[r]},D^{[r]})$ is a vector supermultiplet  
which becomes to the adjoint representation of $U(r)$ gauge symmetry.

As for $\tilde{\mathcal{D}}_s $, $\tilde{\ovl{\mathcal{D}}}_s $ in the lepton sector, we again the same form 
as (\ref{DuRdR}), but the vector supermultiplet meditates $U(2),U(1)$ internal degrees of freedom for $s=l$, $s=e_R$, 
respectively. 

The internal symmetry corresponding to elements of $(M_3(\mathbb{C}),\mathbb{H},\mathbb{C} )$ in the Table 1 amounts to 
$U(3)\times U(2)\times U(1) $. In order to construct unified theory, we relate $U(1)'s $ included in the $U(2),U(3)$ to 
the $U(1)$'s induced by the fluctuations which correspond to $u_a^{[1_j]},j=1,2,3$ in Table 1. 
We choose appropriate $u_a^{1_j}$, $u_{3a}^{(k)},k=0,\ldots,8$ in (\ref{eq2.25c}) and $u_{2a}^{(k)},k=0,\ldots 3$ 
in (\ref{eq2.95}) such that we obtain the following expressions:
\begin{align}
&\sum_{a} c_{a} (\bar{u}_{a}^{[1]})_{\bar{i} \bar{k}} \, \mathcal{D}_{\bar{k} \ell} \, (u_{a}^{[1]})_{\ell j}
\nonumber \\
&=m_1^{-1}\sum_{a} c_{a} (\bar{u}_{a}^{[1_1]})_{\bar{i} \bar{k}} \, \mathcal{D}_{\bar{k} \ell} \, (u_{a}^{[1_1]})_{\ell j} = m_2^{-1}\sum_{a} c_{a} (\bar{u}_{a}^{[1_2]})_{\bar{i} \bar{k}} \, \mathcal{D}_{\bar{k} \ell} \, (u_{a}^{[1_2]})_{\ell j} 
\nonumber \\ 
&=m_3^{-1}\sum_{a} c_{a} (\bar{u}_{a}^{[1_3]})_{\bar{i} \bar{k}} \, \mathcal{D}_{\bar{k} \ell} \, (u_{a}^{[1_3]})_{\ell j} 
\nonumber
\end{align}
\begin{align}
&= z_2^{-1} \sum_{a} c_{a} \left\{(\ovl{u}^{(0)}_{2a})_{\ovl{i}\ovl{k}} \,\mathcal{D}_{\bar{k} \ell} (u^{(0)}_{2a})_{\ell j} + \sum_{m=1,2,3} (\ovl{u}^{(m)}_{2a})_{\ovl{i}\ovl{k}} \,\mathcal{D}_{\bar{k} \ell} (u^{(m)}_{2a})_{\ell j}\right\} 
\nonumber \\
&=  \frac{2}{3} z_3^{-1} \sum_{a} c_{a} \left\{(\ovl{u}^{(0)}_{3a})_{\ovl{i}\ovl{k}} \,\mathcal{D}_{\bar{k} \ell} (u^{(0)}_{3a})_{\ell j} + \sum_{p=1,\cdots,8} (\ovl{u}^{(p)}_{3a})_{\ovl{i}\ovl{k}} \,\mathcal{D}_{\bar{k} \ell} (u^{(p)}_{3a})_{\ell j}\right\} 
\label{eq2.133}
\end{align} 
This operation is to rewrite the notation of the vector supermultiplets as follows: 
\begin{align}
\frac{\lambda_0}{2} \{ G_\mu^{(0)}, \lambda_{3\A}^{(0)}, D_3^{(0)} \} &= z_3 \{ B_\mu, \lambda_{1\A}, D_1 \} \otimes {\bf 1}_3
\label{eq2.134}
\\
\frac{\tau_0}{2} \{ A_\mu^{(0)}, \lambda_{2\A}^{(0)}, D_2^{(0)} \} &= z_2 \{ B_\mu, \lambda_{1\A}, D_1 \} \otimes {\bf 1}_2
\label{eq2.135}
\end{align}
The derived internal symmetry $U(2)$ becomes $SU(2)\otimes U(1)$ and $U(3)$ becomes $SU(3)\otimes U(1)$. 
These $U(1)$'s combined with the other $U(1)$'s shall become to that of weak hypercharge. 

Using (\ref{eq2.134}), we can rewrite (\ref{eq2.36c}) which is the fluctuation of 
$\mathcal{D}$ induced by $u_a^{[3]}$ as follows: 
\begin{align}
V^{[3]}_1 &= z_3 \begin{pmatrix}
0 & 0 & 0 \\
-{i}{\sqrt{2}} \ovl{\lambda}_1^{\dot{\A}} &  \ovl{\sigma}^{\mu\dot{\A}\A} B_\mu & 0 \\
 -D_1 - i\partial^\mu B_\mu  - 2i B_\mu \partial^\mu & -{i}{\sqrt{2}} \lambda_1^\A & 0
\end{pmatrix} \ {\bf 1}_3
\nonumber \\
&+ \sum_{p=1,\cdots,8} \begin{pmatrix}
0 & 0 & 0 \\
-{i}{\sqrt{2}}\, \ovl{\lambda_3}^{(p)\dot{\A}} &  \ovl{\sigma}^{\mu\dot{\A}\A} G_\mu^{(p)} & 0 \\
- D_3^{(p)} - i\partial^\mu G_\mu^{(p)}  - 2i G_\mu^{(p)} \partial^\mu & -{i}{\sqrt{2}} \lambda_3^{(p)\A} & 0
\end{pmatrix} \, \frac{\lambda_p}{2}.
\end{align}
In the same way, using (\ref{eq2.135}), we can rewrite the fluctuation (\ref{eq2.121}) into 
\begin{align}
V^{[2]}_1 &= z_2 \begin{pmatrix}
0 & 0 & 0 \\
-{i}{\sqrt{2}} \ovl{\lambda}_1^{\dot{\A}} & \ovl{\sigma}^{\mu\dot{\A}\A} B_\mu & 0 \\
 -D_1 - i\partial^\mu B_\mu  - 2i B_\mu \partial^\mu & -{i}{\sqrt{2}} \lambda_1^\A & 0
\end{pmatrix} \ {\bf 1}_2
\nonumber \\
&+ \sum_{n=1,2,3} \begin{pmatrix}
0 & 0 & 0 \\
-{i}{\sqrt{2}}\, \ovl{\lambda}_{2}^{(n)\dot{\A}} &  \ovl{\sigma}^{\mu\dot{\A}\A} A_\mu^{(n)} & 0 \\
- D_{2}^{(n)} - i\partial^\mu A_\mu^{(n)}  - 2i A_\mu^{(n)} \partial^\mu & -{i}{\sqrt{2}} \lambda_{2}^{(n)\A} & 0
\end{pmatrix} \ \frac{\tau_n}{2}
\end{align}

Through the above operations, the vector supermultiplet of the $Q$-sector is described as follows:
\begin{align}
\sum_{p=2,3} A_\mu^{[p]} &= \frac{Y}{2} B_\mu + \sum_n \frac{\tau_n}{2} A^{(n)}_\mu + \sum_u \frac{\lambda_u}{2} G^{(u)}_\mu
\label{}
\\
\sum_{p=2,3} \lambda_\A^{[p]} &= \frac{Y}{2} \lambda_{1\A} + \sum_n \frac{\tau_n}{2} \lambda_{2\A}^{(n)} + \sum_u \frac{\lambda_u}{2} \lambda_{3\A}^{(u)}
\label{}
\\
\sum_{p=2,3} D^{[p]} &= \frac{Y}{2} D_1 + \sum_n \frac{\tau_n}{2} D_2^{(n)} + \sum_u \frac{\lambda_u}{2} D_3^{(u)}
\label{}
\end{align}
The vector supermultiplets in the other sector can be written as well. 
The quantum number $Y$ which describes the strength of coupling between 
$\{B_\mu,\lambda_{1\A},D_1 \}$ and quark, lepton is just weak hypercharge. 
The matrix form of weak hypercharge is given by
\begin{align}
\frac{Y}{2} &= \left(
\begin{array}{@{\,}ccccc@{\,}}
z_2+z_3 &  &  & &   \\
 & m_1+z_3 &  &  & \\
 &  & m_2+z_3 & & \\ 
 & & & z_2 &    \\
 &  & &  & m_3 
\end{array}
\right) : \begin{array}{@{\,}c@{\,}}
Q \\
u_R \\
d_R \\
\ell \\
e_R
\end{array}
\label{}
\end{align}
The matrix elements shall be decided such that the electric charge of the particle is given by
\eq
Q = \frac{\tau_3}{2} + \frac{Y}{2},
\label{QYrelation}
\eqend
so that 
\eq
\left\{
\begin{array}{@{\,}l@{\,}}
z_2+z_3 = \frac{1}{6}\\
m_1+z_3 = \frac{2}{3} \\
m_2+z_3 = -\frac{1}{3} \\
z_2 = -\frac{1}{2} \\
m_3 = -1
\end{array}
\right.
\label{}
\eqend
The solution of (\ref{QYrelation}) is given by 
\eq
m_1 = 0, \mbox{ } m_2 = -1, \mbox{ } m_3 = -1, \mbox{ } z_2 = -\frac{1}{2}, \mbox{ } z_3 = \frac{2}{3}.
\label{}
\eqend
Here we can verify ${\rm Tr} \, Y =0$.
\subsection{\normalsize Higgs Supermultiplet}
\ \ \ 
The fluctuation for $\gamma_M\otimes \mathcal{D}_F $ is given by (\ref{fDF}) and (\ref{fDFMd}). 
Taking the commutativity of $M^{(u(d))}$ and $u_a^{[3]}$ into account, non-vanishing elements for the quark sector 
are given by 
\begin{align}
\frac{1}{2} \,\ovl{\mathcal{H}}'_{u(d)} &= \sum_a c'_a \left[\ovl{u}_{a}^{[2]}
M^{(u(d))\dagger} \ovl{u}_a^{[1_{1(2)}]} - M^{(u(d))} \ovl{u}_{a}^{[1_{1(2)}]}\ovl{u}_{a}^{[1_{1(2)}]} \right], \label{Hpu}\\
\frac{1}{2} \,{\mathcal{H}}'_{u(d)} &= \sum_a c'_a 
\left[u_{a}^{[1_{1(2)}]} M^{(u(d))} u_{a}^{[2]} - u_{a}^{[1_{1(2)}]}u_{a}^{[1_{1(2)}]} M^{(u(d))}  \right]. \label{Hpd}
\end{align}
The total Dirac operator modified by the fluctuations which have appeared up to the present in the quark sector is given by
\begin{align}
i\widetilde{\mathcal{D}}_{tot} 
& = \left(
\begin{array}{@{\,}cc|cc@{\,}}
 0  & \widetilde{\ovl{\mathcal{D}}}_Q &  
0 & 0  \\
 \widetilde{\mathcal{D}}_Q & 0 & 
 0 & 
 iy_{s^\prime}^\dagger \ovl{H}^\prime_{s^\prime} \\
 \hline
-iy_{s^\prime} H^\prime_{s^\prime} 
& 0  
& 0  & \widetilde{\ovl{\mathcal{D}}}_s \\
0 & 
0 & 
\widetilde{\mathcal{D}}_s & 0
\end{array}, \right), \label{itildeDtot}
\end{align} 
where $s=u_R(d_R)$, $s^\prime =u(d)$ and 
\begin{align}
{H}'_u &= (\mu_0 \ 0) + \mathcal{H}'_u\ \ H^\prime_d= (0,\mu_0)+\mathcal{H}_d^\prime, \\
\ovl{{H}}'_u & = \begin{pmatrix}
\mu_0 \\
0
\end{pmatrix} + \ovl{\mathcal{H}}'_u ,\ \ 
\ovl{{H}}'_d  = \begin{pmatrix}
0 \\
\mu_0
\end{pmatrix} + \ovl{\mathcal{H}}'_d.
\end{align}
The concrete form of Higgs supermultiplet is given by 
\begin{align}
\ovl{H}'_u &= \begin{pmatrix}
\mu_0 \\
0
\end{pmatrix} + 2\mu_0 \sum_a c'_a \left[\begin{pmatrix}
\bar{u}_{2a}^{(0)}-i\bar{u}_{2a}^{(3)} \\
\bar{u}_{2a}^{(2)}-i\bar{u}_{2a}^{(1)}
\end{pmatrix} \bar{u}_{a}^{[1_1]} -  \begin{pmatrix}
1 \\
0
\end{pmatrix} \bar{u}_{a}^{[1_1]}\bar{u}_{a}^{[1_1]} \right].
\label{eq4.2n}
\end{align} 
Here, putting that  
\eq
1 + 2\sum_a c'_a \left(\bar{u}_{2a}^{(0)} \bar{u}_{a}^{[1_1]} - \bar{u}_{a}^{[1_1]} \bar{u}_{a}^{[1_1]}\right) = v_u^{(0)},
{\rm etc.},
\eqend
(\ref{eq4.2n}) can be expressed by 
\eq
\ovl{H}'_u = \mu_u\begin{pmatrix}
\bar{v}_u^{(0)}-i\bar{v}_u^{(3)} \\
\bar{v}_u^{(2)}-i\bar{v}_u^{(1)}
\end{pmatrix} = \begin{pmatrix}
(H_u^0)^* \\
-(H_u^+)^*
\end{pmatrix},
\label{eq4.4n}
\eqend
so that
\eq
H_u = \begin{pmatrix}
H_u^+ \\
H_u^0
\end{pmatrix}, \mbox{　 } \ovl{H}'_u = (i\tau_2) H_u^*
\eqend
which constructs a isospin doublet.
In the same way, we obtain that 
\eq
H_d = \begin{pmatrix}
H_d^0 \\
H_d^-
\end{pmatrix},
\eqend
where 
\begin{align}
\ovl{H}'_d &= \mu_d\begin{pmatrix}
-\bar{v}_d^{(2)}-i\bar{v}_d^{(1)}, \\
\bar{v}_d^{(0)}+i\bar{v}_d^{(3)}
\end{pmatrix} = \begin{pmatrix}
(H_d^-)^* \\
-(H_d^0)^*
\end{pmatrix} = (i\tau_2) H_d^* ,
\\
H'_u &= \mu_u (v_u^{(0)}+iv_u^{(3)} \ v_u^{(2)}+iv_u^{(1)}) = (H_u^{0} \ -H_u^{+} ) = H_u^T (-i\tau_2),
\\
H'_d &= \mu_d (-v_d^{(2)}+iv_d^{(1)} \ v_d^{(0)}-iv_d^{(3)}) = (H_d^{-} \ -H_d^{0} ) = H_d^T  (-i\tau_2),
\end{align}
and $(H_u^{0})^*$, $(H_u^-)^*$ are expressed by
\eq
(H_u^{0})^* = {\it\Gamma}_0 H_u^{0\dagger} {\it\Gamma}_0,\ (H_u^-)^*={\it\Gamma}_0 H_u^{-\dagger} {\it\Gamma}_0.
\eqend
The elements of $H_u, H_d$ construct chiral multiplets as follows:
\begin{equation}
H_u 
= \begin{pmatrix}
h_u & 0 & 0 \\
\tilde{h}_{u\A} & h_u & 0 \\
F_u & -\tilde{h}_u^\A & h_u
\end{pmatrix},\mbox{ } h_u = \begin{pmatrix}
h_u^+ \\
h_u^0
\end{pmatrix},\ \  
H_d 
= \begin{pmatrix}
h_d & 0 & 0 \\
\tilde{h}_{d\A} & h_d & 0 \\
F_d & -\tilde{h}_d^\A & h_d
\end{pmatrix}, \mbox{ } h_d = \begin{pmatrix}
h_d^0 \\
h_d^-
\end{pmatrix}
\end{equation}
\section{\large Spectral Action}
\ \ \ 
Here, using the modified total Dirac operator (\ref{itildeDtot}), we express 
the action of the kinetic terms of chiral, antichiral  
supermultiplets and their minimum interaction 
with vector supermultiplets including gauge fields and Higgs supermultiplets. 
Let $\Psi_{s+}=(\varphi_s,\psi_{s\A},F_s)$ and $\Psi_{s-}=(\varphi_s^\ast, \bar{\psi}^{\dot{\A}}_s, F_s^\ast )$ 
denote chiral and antichiral 
supermultiplets which represent wave functions of matter particles and their superpartners. 
With the definition of supersymmetric invariant product (\ref{suinvproduct}), the action with regard to 
the chiral and anti-chiral supermultiplets  is given by 
\begin{equation}
I_{matter} = (\Phi,i\widetilde{\mathcal{D}}_{tot}\Phi)=\int_M d^4 x (\mathcal{L}_{kinetic}+\mathcal{L}_{mass}),
\end{equation}
where $\Phi$ is the wave function defined in (\ref{physicalquark}),(\ref{physicallepton}). 
$\mathcal{L}_{kinetic} $, $\mathcal{L}_{mass} $ are expressed,respectively, by
\begin{align}
\lefteqn{
\mathcal{L}_{kinetic}
} \nonumber \\
= & 
\sum_{s=Q,l}\left(\varphi_s^\ast \mathcal{D}_\mu^{(s)}\mathcal{D}^{(s)\mu}\varphi_s
-i\ovl{\psi}_s\ovl{\sigma}^\mu\mathcal{D}_\mu^{(s)}\psi_s+ F_s^\ast F_s
+i\sqrt{2}(\varphi_s^\ast \lambda^{(s)}\psi_s-\ovl{\psi}_s\ovl{\lambda^{(s)}}\varphi_s) 
-\varphi_s^\ast D^{(s)}\varphi_s\right) +
\nonumber \\
\mbox{} &  
\sum_{s=u_R,d_R,e_R}\left(\varphi_s \mathcal{D}_\mu^{(s)}\mathcal{D}^{(s)\mu}\varphi_{s}^\ast
-i\psi_{s}\sigma^\mu\mathcal{D}_\mu^{(s)}\ovl{\psi}_s+ F_s^T F_s^\ast
-i\sqrt{2}(\varphi_s \ovl{\lambda^{(s)}}\ovl{\psi}_s-\psi_s\lambda^{(s)}\varphi_s^\ast)
+\varphi_s D^{(s)}\varphi_s^\ast\right), \label{matterk2}
\end{align}
\begin{align}
\mathcal{L}_{mass} & = \sum_{s=u,d,e} (-i y_s) (\Psi_{s-})^\ast H_s^\prime \Psi_{s+}.
\end{align} 

On the other hand, in the noncommutative geometric approach, 
the action of the vector and Higgs supermultiplets are given by the supersymmetric version of 
the Seeley-DeWitt coefficients of heat kernel expansion of the elliptic operator $P$:
\begin{equation}
Tr_{L^2}f(P) \simeq \sum_{n\geq 0}c_na_n(P),\label{hke}
\end{equation}
where $f(x)$ is an auxiliary smooth function on a smooth compact Riemannian manifold without boundary of 
dimension 4 similar to the non-supersymmetric case.
Since the contribution to $P$ from the antiparticles is 
the same as that of the particles, we consider only the contribution from the particles. Then the 
elliptic operator $P$ in our case is given by the square of the Wick rotated Euclidean Dirac operator 
$\tilde{\mathcal{D}}_{tot}$\cite{paper1}.
Non-vanishing $a_n$'s for $n$ in the flat space are given by
\begin{align}
a_0(P) & =\frac{1}{16\pi^2}\int_M d^4x {\rm tr}_V(\mathbb{I}), \label{a0} \\
a_2(P) & =\frac{1}{16\pi^2}\int_M d^4x {\rm tr}_V(\mathbb{E}), \label{a2} \\
a_4(P) & =\frac{1}{32\pi^2}\int_M d^4x {\rm tr}_V(\mathbb{E}^2 \label{a4}
+\frac{1}{3}\mathbb{E}_{;\mu}^\mu+\frac{1}{6}\Omega_{\mu\nu}\Omega^{\mu\nu}), 
\end{align}
where $\mathbb{E}$ and the bundle curvature $\Omega^{\mu\nu}$ in the flat space are defined as follows:
\begin{align}
\mathbb{E} & = \mathbb{B} -(\partial_\mu \omega^\mu+\omega_\mu\omega^\mu), \label{E}\\
\Omega^{\mu\nu} & = \partial^\mu\omega^\nu -\partial^\nu\omega^\mu+[\omega^\mu,\omega^\nu], \label{Omega}\\
\omega^\mu & =\frac{1}{2}\mathbb{A}^\mu. \label{omega}
\end{align}
We note that the trace for the spin degrees of freedom is the supertrace\cite{paper1} defined by
\begin{align}
{\rm Str} O & = \sum_i\langle i|(-1)^{2s}O|i\rangle \nonumber \\
   & = \sum_b\langle b|O|b\rangle -\sum_f \langle f|O|f\rangle,
\end{align}

The square of total Dirac operator for the quark sector (\ref{itildeDtot}) is given by
\begin{align}
(i{\widetilde{\mathcal{D}}}_{tot})^2 
&= \left(
\begin{array}{@{\,}cc|cc @{\,}}
\widetilde{\ovl{\mathcal{D}}}^{}_Q{\widetilde{{\mathcal{D}}}}^{}_Q & 0 & 0 & D_{14} \\ 
0 & \widetilde{{\mathcal{D}}}^{}_Q\widetilde{\ovl{\mathcal{D}}}^{}_Q & D_{23} & 0 \\ 
\hline
0 & D_{32} & \widetilde{\ovl{\mathcal{D}}}^{}_s \widetilde{{\mathcal{D}}}^{}_s  & 0 \\ 
D_{41} & 0 & 0 & \widetilde{{\mathcal{D}}}^{}_s \widetilde{\ovl{\mathcal{D}}}^{}_s \\
\end{array} \right),
\label{iDtot2}
\end{align}
where 
\begin{align}
D_{14} 
& = i y_{s^\prime}^\dagger \begin{pmatrix}
{F'}_{s^\prime}^* & -{\tilde{h}'^*_{{s^\prime}\dot{\A}}} & {h'}_{s^\prime}^* \\
-i\sqrt{2}\lambda_{\A}^{(Q)}{h'}_{s^\prime}^* + i{\sigma}_{\A\dot{\A}}^\mu \mathcal{D}_\mu^{(Q)} \tilde{h}'^{*\dot{\A}}_{s^\prime} & i{\sigma}^\mu_{\A\dot{\A}} \mathcal{D}_\mu^{(Q)} {h}'^{*}_{s^\prime} & 0 \\
\mathcal{D}^{(Q)}_\mu \mathcal{D}^{(Q)\mu} {h'}_{s^\prime}^* + D^{(Q)}{h'}_{s^\prime}^*  -i\sqrt{2}\ovl{\lambda}_{\dot{\A}}^{(Q)}\tilde{h}'^{*\dot{\A}}_{s^\prime}  
 & -i\sqrt{2} \ovl{\lambda}_{\dot{\A}}^{(Q)}{h'}_u^* & 0 
\end{pmatrix},
\label{D14} 
\end{align}
\begin{align}
D_{41} 
&= -i y_{s^\prime} \begin{pmatrix}
{F'}_{s^\prime} & -\tilde{h}^{'\A}_{{s^\prime}} & {h'}_{s^\prime} \\
 -i\sqrt{2}\ovl{\lambda}^{(s)\dot{\A}}{h'}_{s^\prime} + i{\bar{\sigma}}^{\mu\dot{\A}{\A}} \mathcal{D}_\mu^{(s)} \tilde{h}'_{{s^\prime}{\A}} & i{\bar{\sigma}}^{\mu\dot{\A}{\A}} \mathcal{D}_\mu^{[s]} {h}'_{s^\prime} & 0 \\
\mathcal{D}^{(s)}_\mu \mathcal{D}^{(s)\mu} {h'}_u - D^{(s)}{h'}_{s^\prime} -i\sqrt{2}{\lambda}^{(s)\A}\tilde{h}'_{{s^\prime}\A}  
& -i\sqrt{2} {\lambda}^{[u_R]\A}{h'}_{s^\prime} & 0 
\end{pmatrix},
\label{D41} \\
D_{23} 
&= i y_{s^\prime}^\dagger \begin{pmatrix}
0 & 0 & {h'}_{s^\prime}^* \\
-i\sqrt{2}{h'}_{s^\prime}^* \ovl{\lambda}^{(s)\dot{\A}} & {h'}_{s^\prime}^* (i\partial_\mu + A_\mu^{(s)}){\bar{\sigma}}^{\mu\dot{\A}{\A}} & \tilde{h}'^{*\dot{\A}}_{s^\prime} \\
{h'}_{s^\prime}^* \mathcal{D}^{(s)}_\mu \mathcal{D}^{(s)\mu} - {h'}_{s^\prime}^* D^{(s)}+i\sqrt{2}\tilde{h}'^*_{{s^\prime}\dot{\A}}\ovl{\lambda}^{(s)\dot{\A}}  
& -\tilde{h}'^*_{{s^\prime}\dot{\A}} (i\partial_\mu +A_\mu^{(s)}){\bar{\sigma}}^{\mu\dot{\A}{\A}} & F'^*_{s^\prime} \\
& -i\sqrt{2}{h'}_{s^\prime}^* \lambda^{(s)\A} &     
\end{pmatrix},
\\
D_{32} 
&= -i y_{s^\prime} \begin{pmatrix}
0 & 0 & {h'}_u \\
-i\sqrt{2}{h'}_{s^\prime} {\lambda}^{(Q)}_{\A} & {h'}_s (i\partial_\mu + A_\mu^{(Q)}){{\sigma}}^{\mu}_{\A\dot{\A}} & \tilde{h}'_{s^\prime \A} \\
{h'}_{s^\prime} \mathcal{D}^{(Q)}_\mu \mathcal{D}^{(Q)\mu} + {h'}_s D^{(Q)} +i\sqrt{2}\tilde{h}'^\A_{s^\prime}{\lambda}^{(Q)}_{\A}   
& -\tilde{h}'^{\A}_{s^\prime} (i\partial_\mu + A_\mu^{(Q)}) {{\sigma}}^\mu_{\A\dot{\A}} & F'_{s^\prime} \\   
& -i\sqrt{2}{h'}_{s^\prime} \ovl{\lambda}^{(Q)}_{\dot{\A}} &     
\end{pmatrix}, \label{D32}
\end{align}
where $s =u_R(d_R)$, $s^\prime = u(d)$. 

When we decompose 
\begin{align}
 (i\widetilde{\mathcal{D}}_{tot})^2 &= 
\eta^{\mu\nu} \partial_\mu \partial_\nu \mathbb{I}  + \mathbb{A}^\mu \, \partial_\mu + \mathbb{B},
\label{eq4.31x}
\end{align}
$\omega=\frac{1}{2}\mathbb{A}_\mu$ describes gauge connection and does not have non-diagonal 
elements of ({\ref{iDtot2}), 
because it does not act on the finite space $\mathcal{H}_F$. So the non-diagonal elements which include differential operators 
belong to $\mathbb{B}$. So, we obtain the following expressions:
\begin{align}
\mathbb{E}_{41} & = \mathbb{B} -\partial^\mu\omega_\mu-\omega_\mu\omega^\mu = \mathbb{B}_{41}= D_{41}, \nonumber \\
\mathbb{E}_{32} & = \mathbb{B}_{32} = D_{32},\ 
\mathbb{E}_{23} = \mathbb{B}_{23} = D_{23},\ 
\mathbb{E}_{14} = \mathbb{B}_{14} = D_{14}, 
\end{align} 
We note that the field strength $\Omega^{\mu\nu}$ also does not have non-diagonal element. 

In the previous paper\cite{paper1}, we have already obtained the heat kernel expansion coefficients $a_n(P)$ 
due to $(i\mathcal{D}_M)^2\otimes \bf{1_F}$. 
Let $a_n^{(M)}$ denotes these coefficients. 
$a_0^{(M)}(P)$ and $a_2^{(M)}$ vanish. The action for the gauge fields and its superpartners in 
MSSM comes from  $a_4^{(M)}$ and is given by
\begin{align}
I_{gauge} & =
16\pi^2 f_0 \int_M dx^4 a_4^{(M)}(P) =
 f_0 \int_Mdx^4 \sum_s 
{\rm Tr}\left[
2D^{(s)2} -4\bar{\lambda}_{\dot{\beta}}^{(s)}\bar{\sigma}^{\mu\dot{\beta}\beta}(\mathcal{D}_{\mu}^{(s)}\lambda_{\beta}^{(s)})
 -F_{\mu\nu}^{(s)}F^{(s)\mu\nu}
\right],
\end{align}
where $s$ runs over the elements of $\mathcal{H}_F$, $s=Q,u_R,d_R,l,e_R$ and 
$F^{\mu\nu}_s$ is the field strength of gauge field of the matter particle that $s$ indicates. 

Here, we rescale the vector supermultiplet with 
SU(3) and SU(2) gauge degrees of freedom as follows:
\begin{align}
(A^\mu,\lambda_\A,D) & \rightarrow g_3(A_3^\mu,\lambda_{3\A},D_3)
=\sum_{p=1}^8 g_3 (A_3^{\mu(p)},\lambda_{3\A}^{(p)},D_3^{(p)})\frac{\lambda_p}{2}, \\
(A^\mu,\lambda_\A,D) & \rightarrow g_2(A_2^\mu,\lambda_{2\A},D_2)
=\sum_{p=1}^3 g_2 (A_2^{\mu(p)},\lambda_{2\A}^{(p)},D_2^{(p)}))\frac{\tau_p}{2},
\end{align}
where $\lambda_p$, $\tau_p$ denotes Gell-Mann matrix and Pauli matrix, respectively. 
For the vector supermultiplet with reference to weak hypercharge, we rescale those as 
\begin{equation} 
(B_\mu,\lambda_\A,D)\rightarrow (g_1\frac{Y_s}{2}A_{1\mu},g_1\lambda_{1\A},g_1D_1). 
\end{equation}
The field strength is expressed by
\begin{equation}
F_{j\mu\nu} = \partial_\mu A_{j\nu}-\partial_\nu A_{j\mu}-ig[A_{j\mu},A_{j\nu}].
\end{equation}
Since $A_{\mu}^{(Q)}=g_3A_{3\mu}+g_2A_{2\mu}+g_1\frac{1}{2}\times\frac{1}{3} A_{1\mu}$ and 
${\rm Tr}(\lambda_p\lambda_q)={\rm Tr}(\tau_p\tau_q)=2\delta_{pq}$,
the contribution of $F_{Q\mu\nu}$ to the action is given by
\begin{align}
{\rm Tr}F_{\mu\nu}^{(Q)}F^{(Q)\mu\nu} 
 & = 2\times\frac{1}{2} g_3^2 \sum_j F_{3\mu\nu}^{(j)}F_3^{(j)\mu\nu}
+3\times\frac{1}{2} g_2^2\sum_j F_{2\mu\nu}^{(j)}F_2^{(j)\mu\nu}
 +6\times \frac{1}{4}\left(\frac{1}{3}\right)^2g_1^2F_{1\mu\nu}F_1^{\mu\nu}. \label{FqFq}
\end{align}
In the r.h.s. of (\ref{FqFq}),
the factor $2,3,6$ is from the fact that the left-handed quark $Q$ transforms as (3,2) under the 
gauge group $SU(3)\times SU(2)$. 
In the same way, we obtain that
\begin{align}
\sum_{s=u_R,d_R}{\rm Tr}(F_{\mu\nu}^{(s)}F^{(s)\mu\nu}) & =2\frac{1}{2}g_3^2\sum_j F_{3\mu\nu}^{(j)}F_3^{(j)\mu\nu }+
3\times\frac{1}{4}\left(\left(\frac{4}{3}\right)^2
+\left(-\frac{2}{3}\right)^2\right)g_1^2F_{1\mu\nu}F_1^{\mu\nu} \label{Fud2}\\
\sum_{s=l,e_R}{\rm Tr}(F_{\mu\nu}^{(s)}F^{(s)\mu\nu}) & =\frac{1}{2}g_2^2\sum_j F_{2\mu\nu}^{(j)}F_2^{(j)\mu\nu}+
\frac{1}{4}(2\times (-1)^2+1\times(-2)^2)g_1^2F_{1\mu\nu}F_1^{\mu\nu}. \label{Fle2}
\end{align}
The sum of (\ref{FqFq}),(\ref{Fud2}) and (\ref{Fle2}) amounts to
\begin{equation}
f_0\sum_{s}Tr(F_{\mu\nu}^{(s)}F^{(s)\mu\nu})
=2f_0\left(g_3^2\sum_j F_{3\mu\nu}^{(j)}F_3^{(j)\mu\nu}+g_2^2\sum_j F_{2\mu\nu}^{(j)}F_2^{(j)\mu\nu}
+\frac{5}{3}g_1^2F_{1\mu\nu}F_1^{\mu\nu}\right).
\end{equation}  
Normalizing the Yang-Mills terms to $-\frac{1}{4}F_{n\mu\nu}^{(j)}F_n^{(j)\mu\nu}$ gives:
\begin{equation}
g_3^2=g_2^2=\frac{5}{3}g_1^2,\ \ \ 2f_0g_3^2=\frac{1}{4}. \label{g3g2g1}
\end{equation}
Known already in \cite{connes8}, this expression is same as that of $SU(5)$ grand unification theory.   

The non-diagonal elements of (\ref{iDtot2}) produce Higgs Lagrangian. 
Let us calculate ${\rm tr}_V((\mathbb{E}^2)^{higgs})={\rm Str}(\sum_{i,j}\mathbb{E}_{ij}\mathbb{E}_{ji})$. 
Using (\ref{D14})$\sim $(\ref{D32}), we obtain the result as follows:
\begin{align}
&\| y_s \|^{-2} \mbox{Str} \left( \mathbb{E}_{14} \mathbb{E}_{41} \right) = \| y_s \|^{-2} \mbox{Str} \left(  \mathbb{E}_{23} \mathbb{E}_{32} \right) 
= \| y_s \|^{-2} \mbox{Str} \left( \mathbb{E}_{32} \mathbb{E}_{23}\right) =  \| y_s \|^{-2} \mbox{Str} \left( \mathbb{E}_{41} \mathbb{E}_{14} \right) 
\nonumber \\
&= F'_s F'^*_s - (\partial_\mu h'_s + i h'_s A^{[H_s]}_\mu) (\partial^\mu h'^*_s -i A^{[H_s]\mu} h'^*_s) - i \tilde{h}'^\A_s \sigma^\mu_{\A\dot{\A}} (\partial_\mu \tilde{h}'^{*\dot{\A}}_s -i A^{[H_s]}_\mu \tilde{h}'^{*\dot{\A}}_s)
\nonumber \\
& \hspace{1cm} + i \sqrt{2} \left( \tilde{h}'^{\A}_{s} \lambda^{[H_s]}_{\A} h'^*_s - h'_s \bar{\lambda}^{[H_s]}_{\dot{\A}} \tilde{h}'^{*\dot{\A}}_{s} \right) +  h'_s D^{[H_s]} h'^*_s,
\end{align}
where $s=u(d)$, 
\begin{align}
A_\mu^{[H_u]} & = A_\mu^{(Q)}-A_\mu^{(u_R)},\ 
\lambda_\A^{[H_u]} = \lambda_\A^{(Q)}-\lambda_\A^{(u_R)},\ 
D^{[H_u]}  = D^{(Q)}-D^{(u_R)},\\
A_\mu^{[H_d]} & = A_\mu^{(Q)}-A_\mu^{(d_R)},\ 
\lambda_\A^{[H_d]}  = \lambda_\A^{(Q)}-\lambda_\A^{(d_R)},\ 
D^{[H_d]}  = D^{(Q)}-D^{(d_R)},
\end{align}
and 
\begin{align}
\mbox{diag } y^\dagger_s y_s &= \mbox{diag } y_s y^\dagger_s = \left( |(y_s)_{11}|^2, \, |(y_s)_{22}|^2, \,
 |(y_s)_{33}|^2 \right), \\
\mbox{Tr}\, y^\dagger_s y_s &= \mbox{Tr}\, y_s y^\dagger_s 
\nonumber \\
&= \sum_i |(y_s)_{ii}|^2 \equiv \|y_s\|^2.
\end{align}

Taking ${\rm tr}_V((\mathbb{E}^2)^{higgs})$ due to the lepton sector into account, 
the supersymmetric action of Higgs fields which interact with vector superfields is given by
\begin{align}
I_{Higgs} = & 4f_0\sum_{s=u,d,e}\| y_s \|^2 \left(
F'_s F'^*_s - |\mathcal{D}_\mu^{[H_s]}h^\prime_s|^2
- i \tilde{h}'^\A_s \sigma^\mu_{\A\dot{\A}} \mathcal{D}_\mu^{[H_s]} \tilde{h}'^{*\dot{\A}}_s 
\right. \nonumber \\
& \left.
\hspace{1cm} 
+ i \sqrt{2} \left( \tilde{h}'^{\A}_{s} \lambda^{[H_s]}_{\A} h'^*_s - h'_s \bar{\lambda}^{[H_s]}_{\dot{\A}} \tilde{h}'^{*\dot{\A}}_{s} \right) +  h'_s D^{[H_s]} h'^*_s,
\right) \nonumber \\
= & \frac{1}{8g_3^2}y_u^2\left(
F'_u F'^*_u - |\mathcal{D}_\mu^{[H_u]}h^\prime_u|^2
- i \tilde{h}'^\A_u \sigma^\mu_{\A\dot{\A}} \mathcal{D}_\mu^{[H_u]} \tilde{h}'^{*\dot{\A}}_u 
\right. \nonumber \\
& \left.
\hspace{1cm} 
+ i \sqrt{2} \left( \tilde{h}'^{\A}_{u} \lambda^{[H_u]}_{\A} h'^*_u - h'_u \bar{\lambda}^{[H_u]}_{\dot{\A}} \tilde{h}'^{*\dot{\A}}_{u} \right) +  h'_u D^{[H_u]} h'^*_u,
\right) \nonumber \\
& + \frac{1}{8g_3^2}(y_d^2+y_e^2)\left(
F'_d F'^*_d - |\mathcal{D}_\mu^{[H_d]}h^\prime_d|^2
- i \tilde{h}'^\A_d \sigma^\mu_{\A\dot{\A}} \mathcal{D}_\mu^{[H_d]} \tilde{h}'^{*\dot{\A}}_d 
\right. \nonumber \\
& \left.
\hspace{1cm} 
+ i \sqrt{2} \left( \tilde{h}'^{\A}_{d} \lambda^{[H_d]}_{\A} h'^*_d - h'_d \bar{\lambda}^{[H_d]}_{\dot{\A}} \tilde{h}'^{*\dot{\A}}_{d} \right) +  h'_d D^{[H_d]} h'^*_d,
\right),
\end{align}
where we use (\ref{g3g2g1}). Then we have obtained all the terms of correct Lagrangian which give MSSM.  
\section{\large CONCLUSIONS}
\ \ \ 
We defined the "triple" extended from the spectral triple which was to specify NCG. As the functional space,
we take chiral and antichiral supermultiplets which correspond to matter fields and 
their superpartners in the supersymmetric standard model.  
we introduced algebra $\mathcal{A}$ and total Dirac operator 
$i\mathcal{D}_{tot}=i\mathcal{D}_M\otimes 1+\gamma_M\otimes \mathcal{D}_F $ in the flat space-time 
which acted on the functional space $\mathcal{H}=\mathcal{H}_M\otimes \mathcal{H}_F$. 
We considered the internal fluctuations induced by the elements of $\mathcal{A}$. The fluctuation for 
$i\mathcal{D}_M \otimes 1$ generated vector supermultiplets. 
The vector supermultiplets meditate $U(3)\otimes U(2)\otimes U(1)$ gauge degrees of freedom. 
We took out $U(1)$'s from $U(3)$, $U(2)$ and combined the other $U(1)$'s to obtain $U(1)$ of weak hypercharge so that 
the gauge degrees of freedom amounted to those of MSSM, $SU(3)\times SU(2)\times U(1)_Y$ and 
each matter particle was distributed to adequate quantum numbers.   
On the other hand, the fluctuation for $\gamma_M\otimes \mathcal{D}_F $ generated supermultiplets 
which transformed as Higgs fields of MSSM.   

Following the 
supersymmetric version of spectral action principle, we calculated the square of the total Dirac operator 
and Seeley-Dewitt coefficients of heat kernel expansion. From the coefficient $a_4(P)^{(M)}$, we obtained 
the action of vector supermultiplets of MSSM. Normalizing the coefficients of 
the squared field strength of each of $SU(3)$,$SU(2)$,$U(1)_Y$ gauge field to the same value, we found the relations 
between coupling constants which was same as that of SU(5) grand unification theory. 
We also verified that 
the coefficient due to non-diagonal elements of total Dirac operator gave the action for the 
Higgs supermultiplets and that we arrived at the correct whole action of MSSM. 

All formulae in this paper were established in the Minkowskian space. 
We are now preparing the theory in which 
the total Dirac operator in the curved Riemannian space is assumed in order to take gravity into account. 
It will give the supersymmetric version of NC geometric view of unifying the gravity and gauge,Higgs fields. 
\newpage
\appendix
{\bf \Large Appendix }
\renewcommand{\theequation}{A.\arabic{equation}}
\setcounter{equation}{0}
\section{Vector supermultiplet with U(r) gauge symmetry}
\ \ \ 
In this appendix, we will see that choosing suitable components of (\ref{Ua}),(\ref{barUa}), we can construct 
the vector supermultiplet in (\ref{Amu})$\sim$(\ref{D}) which meditates the $U(r)$ gauge degrees of freedom.
\subsection{\large U(1)}
\ \ \ 
Choosing as the algebra the space of complex functions, $u_a^{[1_j]},j=1,2,3$ in (\ref{Ua}),(\ref{barUa}), the vector supermultiplet defined in (\ref{Amu})$\sim $(\ref{D}) meditate U(1) gauge symmetry of which the 
phase $m_j(j=1,2,3)$ in the Table 1 represents the quantum number of $u_R,d_R$,$e_R$ field, respectively. 
%
%
%
\subsection{\large U(3)}
\ \ \ 
In this case, the elements of $\mathcal{A}_F \otimes \mathcal{A}_M$ are 
$3 \times 3$ matrix-valued functions ${(u_{a}^{[3]}(x))_A^{\ B}}, A,B = 1,2,3$ which act on the internal degrees of freedom 
of quarks $A=1,2,3$. The $3 \times 3$ matrix-valued functions ${(u_{a}^{[3]})_A^{\ B}} \in \mathcal{A}_F \otimes \mathcal{A}_+$
 and ${(\ovl{u}_{a}^{[3]})_A^{\ B}} \in \mathcal{A}_F \otimes \mathcal{A}_-$ can be expressed in the matrix form as follows:
\begin{align}
{(u_{a}^{[3]})_A^{\ B}} &= \sum_{k=0,1,\cdots,8} u_{3a}^{(k)}(x)\, (\frac{\lambda_k}{2})_A^{\ B}
\label{eq2.25c}
\\
{(\ovl{u}_{a}^{[3]})_A^{\ B}} &= \sum_{k=0,1,\cdots,8} \ovl{u}_{3a}^{(k)}(x)\, (\frac{\lambda_k}{2})_A^{\ B}
\label{eq2.26c}
\end{align}
where $\lambda_k$ are Gell-Mann matrices.

The internal fluctuation due to $(u_a^{[3]})_A^{\ B}$ and $(u_{ab}^{[33]})_{AC}^{\ BD} ={(u_{a}^{[3]})_A^{\ B}} {(u_{a}^{[3]})_C^{\ D}}$ is given by
\begin{align}
(V^{[3]}_{WZ})_{\bar{i}j\, A}^{\ \ \ C} &
&= 2\sum_{a} c_{a}^{} [(\bar{u}_a^{[3]})_{\bar{i} \bar{k}}]_A^{\ A'} \, \mathcal{D}_{\bar{k} \ell} \, [(u_a^{[3]})_{\ell j}]_{A'}^C 
+ 2\sum_{a,b} c_{a} c_b [(\bar{u}_{ab}^{[33]})_{\bar{i} \bar{k}}]_{AB}^{\ A'B'}  \, \mathcal{D}_{\bar{k} \ell} \, [(u_{ab}^{[33]})_{\ell j}]_{A'B'}^{\ BC} 
\label{eq2.27c}
\end{align}
Substituting (\ref{eq2.25c}), (\ref{eq2.26c}), the fluctuation due to the part of 
$({u_a^{[3]}})_A^{\ B}$ is obtained by 
\begin{align}
(V_{\bar{i}j}^{[3]})_A^{\ C} &= 2 \sum_{a} c_{a} {[(\bar{u}^{[3]}_{a})_{\bar{i} \bar{k}}]_A}^B \, \mathcal{D}_{\bar{k} \ell} \, {[(u_{a}^{[3]})_{\ell j}]_B}^C 
\nonumber \\
&= 2 \sum_{a} c_{a} \left\{ \sum_{p=0,\cdots,8} (\ovl{u}^{(p)}_{3a})_{\ovl{i}\ovl{k}} \,\mathcal{D}_{\bar{k} \ell} (u^{(p)}_{3a})_{\ell j} \frac{1}{\sqrt{6}} (\frac{\lambda_0}{2})_A^{\ C} + \sum_u\left[(\ovl{u}^{(0)}_{3a})_{\ovl{i}\ovl{k}} \,\mathcal{D}_{\bar{k} \ell} (u^{(u)}_{3a})_{\ell j} \frac{1}{\sqrt{6}} \right.\right.
\nonumber \\
& \hspace{0cm} \left.\left. +(\ovl{u}^{(u)}_{3a})_{\ovl{i}\ovl{k}} \,\mathcal{D}_{\bar{k} \ell} (u^{(0)}_{3a})_{\ell j} \frac{1}{\sqrt{6}} + \sum_{p,t,=1,\cdots,8} \frac{1}{2}(\ovl{u}^{(p)}_{3a})_{\ovl{i}\ovl{k}} \,\mathcal{D}_{\bar{k} \ell} (u^{(t)}_{3a})_{\ell j} \,(i f_{ptu}+ d_{ptu}) \right] (\frac{\lambda_u}{2})_A^{\ C} \right\},
\label{eq2.28c}
\end{align}
where we use the following formulae:
\begin{align}
\lambda_0 &= 
\sqrt{2/3} \, \, \textrm{diag}(1,1,1)
 \label{eq2.29c}
 \\
 [\lambda_p, \lambda_t] &= 2i f_{ptu} \lambda_u, \mbox{ } 
 \{\lambda_p, \lambda_t \} = 2 d_{ptu} \lambda_u + \frac{4}{3} \delta_{pt}
 \mbox{ } (p,t,u = 1, \cdots, 8)
 \label{eq2.30c}
 \end{align}
From (\ref{ua}),(\ref{barua}),(\ref{DM}), the product of $\bar{u}_a$,$\mathcal{D}$,$u_a $ is given by 
\begin{align}
\bar{u}_{a\bar{i} \bar{k}} \, \mathcal{D}_{\bar{k} \ell} \, u_{a\ell j}  
&= \frac{1}{m_0^2}  \begin{pmatrix}
\varphi_a^* & 0 & 0 \\
\ovl{\psi}_a^{\dot{\A}} & \varphi_a^* & 0 \\
F_a^* & -\ovl{\psi}_{a\dot{\A}} & \varphi_a^*
\end{pmatrix}_{\bar{i} \bar{k}} \begin{pmatrix}
0 & 0 & 1 \\
0 & i\ovl{\sigma}^\mu \partial_\mu & 0 \\
\square & 0 & 0
\end{pmatrix}_{\bar{k} \ell} 
\begin{pmatrix} 
\varphi_a & 0 & 0 \\
\psi_{ a \A} & \varphi_a & 0 \\
F_a & -\psi_a^\A & \varphi_a
\end{pmatrix}_{\ell j}
\nonumber \\
& = \frac{1}{m_0^2}  \begin{pmatrix}
\varphi_a^* F_a & -\varphi_a^* \psi_a^\A & \varphi_a^* \varphi_a \\
\ovl{\psi}^{a\dot{\A}} F_a + i \varphi_a^* \ovl{\sigma}^{\mu\dot{\A}\A} \partial_\mu \psi_{a\A} & -\ovl{\psi}_a^{\dot{\A}} \psi_a^\A + i \ovl{\sigma}^{\mu\dot{\A}\A} \varphi_a^* \partial_\mu \varphi_a & \ovl{\psi}_a^{\dot{\A}} \varphi \\
F_a^* F_a -i \ovl{\psi}_{a\dot{\A}} \ovl{\sigma}^{\mu\dot{\A}\A} \partial_\mu \psi_{{a\A}} + \varphi_a^* \square \varphi_a & -F_a^* \psi_a^\A -i \ovl{\psi}_{a\dot{\A}} \ovl{\sigma}^{\mu\dot{\A}\A} \partial_\mu \varphi_a & F_a^* \varphi_a
\end{pmatrix}
\label{eq2.10c}
\end{align}
In the way same as (\ref{eq2.10c}), we obtain the following formula:
\begin{align}
\sum_a c_a (\ovl{u}^{(p)}_{3a})_{\ovl{i}\ovl{k}} \,\mathcal{D}_{\bar{k} \ell} (u^{(t)}_{3a})_{\ell j} &= \frac{1}{2} \begin{pmatrix}
0 & 0 & 0 \\
-{i}{\sqrt{2}} \ovl{\lambda_3}^{(p,t)\dot{\A}} &  \ovl{\sigma}^{\mu\dot{\A}\A} G_\mu^{(p,t)} & 0 \\
- D_3^{(p,t)} - i\partial^\mu G_\mu^{(p,t)}  - 2i G_\mu^{(p,t)} \partial^\mu & -{i}{\sqrt{2}} \lambda_3^{(p,t)\A} & 0
\end{pmatrix}
\label{eq2.31c}
\\
G_\mu^{(p,t)} &= \frac{i}{m_0^2} \sum_{a} c_{a} \left( \varphi_{3a}^{(p) *} \partial_\mu \varphi_{3a}^{(t)} - \partial_\mu \varphi_{3a}^{(p) *} \varphi_{3a}^{(t)} -i\, \ovl{\psi}_{3a}^{(p)} \ovl{\sigma}^\mu \psi_{3a}^{(t)} \right)
\label{eq2.32c}
\\
\lambda_{3\A}^{(p,t)} &= -\sqrt{2} \frac{i}{m_0^2}\sum_{a} c_{a} \left(
F_{3a}^{(p)*} \psi^{(t)}_{ {3a}\A} - i \sigma^\mu_{\A\dot{\A}} \ovl{\psi}_{3a}^{(p)\dot{\A}*} \partial_\mu \varphi^{(t)}_{3a} \right)
\label{eq2.33c}\\
D_3^{(p,t)} &= -\frac{1}{m_0^2}  
\sum_{a} c_{a} \left[ 2  F_{3a}^{(p)*} F_{3a}^{(t)}-2(\partial^\mu \varphi_{3a}^{(p)*} \partial_\mu \varphi_{3a}^{(t)})  \right.
\nonumber \\
&\left. \hspace{3cm} + i  \left\{ \partial_\mu \ovl{\psi}^{(p)}_{{3a}\dot{\A}} \ovl{\sigma}^{\mu \dot{\A}\A} \psi^{(t)}_{{3a}\A} -\ovl{\psi}^{(p)}_{{3a}\dot{\A}} \ovl{\sigma}^{\mu \dot{\A}\A} \partial_\mu \psi^{(t)}_{{3a}\A} \right\}  \right],
\label{eq2.34c}  
\end{align}
where we use the Wess-Zumino condition
\eq
\sum_a c_a {(\varphi_{a}^{[3]*} \varphi_{a}^{[3]})_A}^C = \sum_a c_a{(\varphi_{a}^{[3]*} \psi_{a}^{[3]\A})_A}^C = \sum_a c_a {(\varphi_{a}^{[3]*} F_{a}^{[3]})_A}^C = 0.
\eqend
Substituting (\ref{eq2.31c}) into (\ref{eq2.28c})，we obtain the following formulae:
\begin{align}
(V_{\bar{i}j}^{[3]})_A^{\ C}
& = 2 \sum_{a} c_{a}^{} {[(\bar{u}_{a}^{[3]})_{\bar{i} \bar{k}}]_A^{\ B}} \, \mathcal{D}_{\bar{k} \ell} \ {[(u_{a}^{[3]})_{\ell j}]_B^{\ C}} 
\nonumber \\ 
&  = \sum_{p=0,\cdots,8} \begin{pmatrix}
0 & 0 & 0 \\
-{i}{\sqrt{2}} \ovl{\lambda_3}^{(p)\dot{\A}} &  \ovl{\sigma}^{\mu\dot{\A}\A} G_\mu^{(p)} & 0 \\
- D_3^{(p)} - i\partial^\mu G_\mu^{(p)}  - 2i G_\mu^{(p)} \partial^\mu & -{i}{\sqrt{2}} \lambda_3^{(p)\A} & 0
\end{pmatrix} \, \left(\frac{\lambda_p}{2}\right)_A^{\ C},
\label{eq2.36c}
\end{align}
where 
\begin{align}
G_\mu^{(0)} &= \frac{1}{\sqrt{6}} \sum_{p=0,\cdots,8} G_\mu^{(p,p)}
\label{eq2.37c}
\\
\lambda_3^{(0)\A} &= \frac{1}{\sqrt{6}}\sum_{p=0,\cdots,8} \lambda_3^{(p,p)\A}
\label{eq2.38c}\\
D_3^{(0)} &= \frac{1}{\sqrt{6}} \sum_{p=0,\cdots,8}  D_3^{(p,p)}
\label{eq2.39c}
\end{align}
and 
\begin{align}
G_\mu^{(u)} &= \frac{1}{\sqrt{6}} G_\mu^{(0,u)} + \frac{1}{\sqrt{6}} G_\mu^{(u,0)} + \frac{1}{2} \sum_{p,t=1,\cdots,8} (i f_{ ptu} + d_{ptu})  G_\mu^{(p,t)}
\label{eq2.40c}
\\
\lambda_3^{(u)\A} &= \frac{1}{\sqrt{6}}\lambda_3^{(0,u)\A}
 + \frac{1}{\sqrt{6}}\lambda_3^{(u,0)\A}
 + \frac{1}{2} \sum_{p,t=1,\cdots,8} (i f_{ptu} + d_{ptu}) \lambda_3^{(p,t)\A}
\label{eq2.41c}
\\
D_3^{(u)} &= \frac{1}{\sqrt{6}}  D_3^{(0,u)}
 + \frac{1}{\sqrt{6}}  D_3^{(u,0)}
 + \frac{1}{2} \sum_{p,t=1,\cdots,8} (i f_{ptu} + d_{ptu}) D_3^{(p,t)}.
\label{eq2.42c}
\end{align}
In (\ref{eq2.32c}) and (\ref{eq2.34c}),
\begin{align}
G_\mu^{(p,t)*} &= G_\mu^{(t,p)}
\\
D_3^{(p,t)*} &= D_3^{(t,p)},
\end{align}
so that $G_\mu^{(u)}$ and $D_3^{(u)}$ are real functions, while $\lambda_3^{(u)\A}$ is a complex spinor field. 
The vector supermultiplet ($G_\mu^{(u)}$, $\lambda_3^{(u)\A}$, $D_3^{(u)}$), $u=0,1,\cdots,8$ 
obeys the transformation law of the adjoint representation of  $U(3)$. 
(\ref{eq2.28c}) is equivalent to decomposing the tensor product $\bm{8} \otimes \bm{8}$ under $SU(3) \subset U(3)$ into 
the direct sum of irreducible representations and extract the term of $\bm{1} \oplus \bm{8}$.

Next，we deal with the fluctuation due to the part of
${u_{ab}^{[33]}}_{AC}^{BD} ={(u_{a}^{[3]})_A^{\ B}} {(u_{a}^{[3]})_C^{\ D}}$. 
As for the representation of $SU(3)$，it is to extract $(\bm{1} \oplus \bm{8}) \times (\bm{1} \oplus \bm{8})$ 
from irreducible representations into which the tensor product $(\bm{8} \otimes \bm{8}) \otimes (\bm{8} \otimes \bm{8})$ 
is decomposed, where $\times $ denote the product of representation matrix.
%
So we contract the tensor product as shown in the second term of (\ref{eq2.27c}). 
We let $V_{\bar{i}j}^{\prime[33]} $ denote the second term of (\ref{eq2.27c}). 
Under the Wess-Zumino condition, we calculate the matrix elements of $V'^{[33]}_{\bar{i}j}$:
\begin{align}
\frac{1}{2} (V'^{[33]}_{\bar{1} 1})_A^{\ C} &
= \frac{1}{m_0^2} \sum c_a c_b (\varphi_{ab}^{[33]*})_{AB}^{A'B'} (F_{ab}^{[33]})_{A'B'}^{BC}
\nonumber \\
&= \frac{1}{m_0^4} \sum c_a c_b (\varphi_a^{[3]*})_A^{A'} (\varphi_b^{[3]*})_B^{B'} \left[(\varphi_a^{[3]})_{A'}^{B} (F_b^{[3]})_{B'}^C + (F_a^{[3]})_{A'}^{B}(\varphi_b^{[3]})_{B'}^C - (\psi_a^{[3]})_{A'}^{B}(\psi_b^{[3]})_{B'}^C \right]
\nonumber \\
&= \frac{1}{m_0^4} \sum c_a c_b \left[(\varphi_a^{[3]*}\varphi_a^{[3]})_{A}^{B}(\varphi_b^{[3]*} F_b^{[3]})_{B}^C + (\varphi_a^{[3]*} F_a^{[3]})_{A}^{B}(\varphi_b^{[3]*} \varphi_b^{[3]})_{B}^C - (\varphi_a^{[3]*} \psi_a^{[3]})_{A}^{B}(\varphi_b^{[3]*} \psi_b^{[3]})_{B}^C \right]
\nonumber \\
&= 0.
\label{eq2.44c}
\end{align}
Continuing the similar calculations, we see that the matrix elements vanish except for $V'^{[33]}_{\bar{3}1}$.

The element $V'^{[33]}_{\bar{3}1}$ is given as follows:
\begin{align}
\frac{1}{2} (V'^{[33]}_{\bar{3} 1})_A^{\ C} &= \sum_{a,b} c_a c_b \,[(\ovl{u}_{ab}^{[33]})_{\bar{3}\bar{k}}]_{AB}^{A'B'} \mathcal{D}_{\bar{k} \ell} \ [(u_{ab}^{[33]})_{\ell 1}]_{A'B'}^{BC} 
\nonumber \\
&= \frac{1}{m_0^2} \sum c_a c_b \left[(F_{ab}^{[33]*})_{AB}^{A'B'} (F_{ab}^{[33]})_{A'B'}^{BC} -i (\ovl{\psi}_{ab}^{[33]*})_{AB}^{A'B'} \ovl{\sigma}^\mu \partial_\mu (\psi_{ab}^{[33]})_{A'B'}^{BC} \right. 
\nonumber \\
& \left. \hspace{8cm} + (\varphi_{ab}^{[33]*})_{AB}^{A'B'} \square (\varphi_{ab}^{[33]})_{A'B'}^{BC}
\right] \label{V3331s}
\\
& m_0^2 (F_{ab}^{[33]*})_{AB}^{A'B'} (F_{ab}^{[33]})_{A'B'}^{BC} = (\ovl{\psi}_a^{[3]} \psi_a^{[3]})_A^B \,(\ovl{\psi}_b^{[3]} \psi_b^{[3]})_B^C + \mbox{(terms which vanish by W-Z condition)}
\nonumber \\
&\hspace{3.9cm} = -\frac{1}{2} (\ovl{\psi}_a^{[3]} \bar{\sigma}^\mu \psi_a^{[3]} )_A^B \, (\ovl{\psi}_b^{[3]} \bar{\sigma}_\mu \psi_b^{[3]} )_B^C
\label{eq2.46c} 
\end{align}
\begin{align}
& m_0^2 (\ovl{\psi}_{ab}^{[33]*})_{AB}^{A'B'} \ovl{\sigma}^\mu \partial_\mu (\psi_{ab}^{[33]})_{A'B'}^{BC} = (\ovl{\psi}_a^{[3]} \bar{\sigma}^\mu \psi_a^{[3]})_A^B \, (\varphi_b^{[3]*} \partial_\mu \varphi_b^{[3]})_B^C + (\varphi_a^{[3]*} \partial_\mu \varphi_a^{[3]})_A^{B} \, (\ovl{\psi}_b^{[3]} \bar{\sigma}^\mu\psi_b^{[3]})_B^C,
\nonumber \\
&\hspace{5.75cm} + \mbox{(terms which vanish by W-Z condition)},
\\
& m_0^2 (\varphi_{ab}^{[33]*})_{AB}^{A'B'} \square (\varphi_{ab}^{[33]})_{A'B'}^{BC} = (\varphi_a^{[3]*})_A^{A'} (\varphi_b^{[3]*})_B^{B'} \partial^\mu \partial_\mu (\varphi_a^{[3]})_{A'}^B (\varphi_b^{[3]})_{B'}^C
\nonumber \\
&\hspace{4.1cm} = 2 \,(\varphi_a^{[3]*} \partial_\mu \varphi_a^{[3]})_A^B \, (\varphi_b^{[3]*} \partial^\mu \varphi_b^{[3]})_B^C
\nonumber \\
&\hspace{5.75cm} + \mbox{(terms which vanish by W-Z condition)},
\end{align}
so that 
\begin{align}
\frac{1}{2} (V'^{[33]}_{\bar{3} 1})_A^{\ C} &= \frac{1}{m_0^4} \sum_{a,b} c_a c_b \,\left[ -\frac{1}{2} (\ovl{\psi}_a^{[3]} \bar{\sigma}^\mu \psi_a^{[3]} )_A^B \, (\ovl{\psi}_b^{[3]} \bar{\sigma}_\mu \psi_b^{[3]} )_B^C -i \,\ovl{\psi}_a^{[3]} \bar{\sigma}^\mu \psi_a^{[3]})_A^B \, (\varphi_b^{[3]*} \partial_\mu \varphi_b^{[3]})_B^C 
\right.
\nonumber \\
& \left. \hspace{2.5cm} -i \, (\varphi_a^{[3]*} \partial_\mu \varphi_a^{[3]})_A^{B} \, (\ovl{\psi}_b^{[3]} \bar{\sigma}^\mu\psi_b^{[3]})_B^C + 2 \,(\varphi_a^{[3]*} \partial_\mu \varphi_a^{[3]})_A^B \, (\varphi_b^{[3]*} \partial^\mu \varphi_b^{[3]})_B^C 
\right]
\nonumber \\
&= - \frac{1}{2} (A_\mu^{[3]} A^{[3] \mu})_A^{\ C}.
\label{eq2.50c}
\end{align}
Here, using (\ref{eq2.37c}) and (\ref{eq2.40c}), $A_\mu^{[3]}$ is given by 
\begin{align}
A_\mu^{[3]} &= \frac{i}{m_0^2} \sum_a c_a \left[\left(\varphi_a^{[3]*}\partial_\mu \varphi_a^{[3]} - \partial_\mu \varphi_a^{[3]*}\varphi_a^{[3]} \right) - i \ovl{\psi}^{[3]}_{a\dot{\A}}\ovl{\sigma}_\mu^{\dot{\A}\A}\psi^{[3]}_{a\A} \right] 
\nonumber \\
&= \frac{i}{m_0^2} \sum_a c_a \left[2 \,\varphi_a^{[3]*}\partial_\mu \varphi_a^{[3]}  - i \ovl{\psi}^{[3]}_{a\dot{\A}}\ovl{\sigma}_\mu^{\dot{\A}\A}\psi^{[3]}_{a\A} \right]
\nonumber \\
&=\sum_{p=0,\cdots,8} G_\mu^{(p)} \frac{\lambda_p}{2}.
\label{eq2.51c}
\end{align}，

The internal fluctuation (\ref{eq2.27c}) due to (\ref{eq2.36c}), (\ref{eq2.50c}) amounts to 
\begin{align}
(V^{[3]}_{WZ})_{\bar{i}j} &= V^{[3]}_{\bar{i}j} + V'^{[33]}_{\bar{i}j}
\nonumber \\
&= 
\begin{pmatrix}
0 & 0 & 0 \\
-{i}{\sqrt{2}} \ovl{\lambda}^{[3]\dot{\A}} &  \ovl{\sigma}^{\mu\dot{\A}\A} A_\mu^{[3]} & 0 \\
- D^{[3]} - i\partial^\mu A_\mu^{[3]}  - 2i A_\mu^{[3]} \partial^\mu - A_\mu^{[3]}A^{[3]\mu} & -{i}{\sqrt{2}} \lambda^{[3]\A} & 0
\end{pmatrix}, \label{V3WZ}
\end{align}
where 
\begin{align}
\lambda^{[3]}_\A &= \sum_{p=0,\cdots,8} \lambda_{3\A}^{(p)} \frac{\lambda_p}{2}, 
\\
D^{[3]} &= \sum_{p=0,\cdots,8} D_3^{(p)} \frac{\lambda_p}{2}. 
\end{align}

Repeating the similar calculation, the fluctuation for the opposite chirality sector is given by 
\begin{align}
(\ovl{V}_{WZ}^{[3]})_{{i}\bar{j}} 
& =
\begin{pmatrix}
0 & 0 & 0 \\
-{i}{\sqrt{2}} {\lambda}_{{\A}}^{[3]} &  {\sigma}^{\mu}_{\A\dot{\A}} A_\mu ^{[3]} & 0 \\
 D^{[3]} - i\partial^\mu A_\mu^{[3]}  - 2i A_\mu^{[3]} \partial^\mu - A_\mu^{[3]} A^{{[3]}\mu} & -{i}{\sqrt{2}} \ovl{\lambda}_{\dot{\A}}^{[3]} & 0
\end{pmatrix}. \label{barVWZ3}
\end{align}


%
%
%
\subsection{\large U(2)}
\ \ \ 
Here, we calculate internal fluctuation due to quaternion-valued function.
Quaternion-valued chiral multiplet ${(u_a^{[2]})_I}^J \in \mathcal{A}_F \otimes \mathcal{A}_+$ is expressed into matrix 
form as follows:
\begin{align}
{(u_{a}^{[2]})_I}^J &= {\begin{pmatrix}
u_{2a}^{(0)}(x) + i u_{2a}^{(3)}(x) & u_{2a}^{(2)}(x) + i u_{2a}^{(1)}(x) \\
-u_{2a}^{(2)}(x) + i u_{2a}^{(1)}(x) & u_{2a}^{(0)}(x) - i u_{2a}^{(3)}(x)
\end{pmatrix}_I}^J
\nonumber \\
&= {(\tau_0)_I}^J u_{2a}^{(0)}(x)  + {(i\tau_1)_I}^J u_{2a}^{(1)}(x) + {(i\tau_2)_I}^J u_{2a}^{(2)}(x) + {(i\tau_3)_I}^J u_{2a}^{(3)}(x),
\label{eq2.93}
\end{align}
where 
$u_{2a}^{(k)}(x), (k=0,1,2,3)$ is given by
\eq
(u_{2a}^{(k)})_{ij} = \frac{1}{m_0} \begin{pmatrix}
\varphi_{2a}^{(k)}  & 0 & 0 \\
\psi_{2a\A}^{(k)} & \varphi_{2a}^{(k)}  & 0 \\
F_{2a}^{(k)} & -\psi_{2a}^{(k)\A} & \varphi_{2a}^{(k)} 
\end{pmatrix}.
\label{eq2.94}
\eqend
The matrix form of 	
$(\ovl{u}_{a}^{[2]})_{IJ} \in \mathcal{A}_F \otimes \mathcal{A}_-$ is also given by
\begin{align}
{(\ovl{u}_{a}^{[2]})_I}^J &= {\begin{pmatrix}
\ovl{u}_{2a}^{(0)}(x) - i \ovl{u}_{2a}^{(3)}(x) & -\ovl{u}_{2a}^{(2)}(x) - i \ovl{u}_{2a}^{(1)}(x) \\
\ovl{u}_{2a}^{(2)}(x) - i \ovl{u}_{2a}^{(1)}(x) & \ovl{u}_{2a}^{(0)}(x) + i \ovl{u}_{2a}^{(3)}(x)
\end{pmatrix}_I}^J
\nonumber \\
&= {(\tau_0)_I}^J \ovl{u}_{2a}^{(0)}(x)  - {(i\tau_1)_I}^J \ovl{u}_{2a}^{(1)}(x) - {(i\tau_2)_I}^J \ovl{u}_{2a}^{(2)}(x) - {(i\tau_3)_I}^J \ovl{u}_{2a}^{(3)}(x),
\label{eq2.95}
\end{align}
\begin{align}
(\ovl{u}_{2a}^{(k)})_{ij} &= \frac{1}{m_0} \begin{pmatrix}
\varphi_{2a}^{(k)*}  & 0 & 0 \\
\ovl{\psi}_{2a}^{(k)\dot{\A}} & \varphi_{2a}^{(k)*}  & 0 \\
F_{2a}^{(k)*} & -\ovl{\psi}_{2a\dot{\A}}^{(k)} & \varphi_{2a}^{(k)*} 
\end{pmatrix}.
\label{eq2.96}
\end{align}
Developing 
$\bar{u}_a^{[2]} \mathcal{D}^{(2)} u^{[2]}_a$ with the aid of (\ref{eq2.93}), (\ref{eq2.95}), we obtain 
\begin{align}
&\sum_{a} c_{a}^{} [(\bar{u}_{a}^{[2]})_{\bar{i} \bar{k}}]_I^{\ I'} \, \mathcal{D}_{\bar{k} \ell} \, [(u_{a}^{[2]})_{\ell j}]_{I'}^{\ J} 
\nonumber \\
&= 
\sum_{a} c_{a}^{} \left[\left\{(\ovl{u}^{(0)}_{2a})_{\ovl{i}\ovl{k}} \,\mathcal{D}_{\bar{k} \ell} (u^{(0)}_{2a})_{\ell j} + \sum_{m=1,2,3} (\ovl{u}^{(m)}_{2a})_{\ovl{i}\ovl{k}} \,\mathcal{D}_{\bar{k} \ell} (u^{(m)}_{2a})_{\ell j}\right\} (\tau_0)_I^{\ J} \right.
\nonumber \\
&\hspace{1cm}\left. + i\sum_{n=1,2,3}\left\{ (\ovl{u}^{(0)}_{2a})_{\ovl{i}\ovl{k}} \,\mathcal{D}_{\bar{k} \ell} (u^{(n)}_{2a})_{\ell j} - (\ovl{u}^{(n)}_{2a})_{\ovl{i}\ovl{k}} \,\mathcal{D}_{\bar{k} \ell} (u^{(0)}_{2a})_{\ell j} \right\}(\tau_n)_I^{\ J}   \right.
\nonumber \\
&\hspace{1cm}\left. + \frac{i}{2}\sum_{m,s,n=1,2,3}\left\{ (\ovl{u}^{(m)}_{2a})_{\ovl{i}\ovl{k}} \,\mathcal{D}_{\bar{k} \ell} (u^{(s)}_{2a})_{\ell j} - (\ovl{u}^{(s)}_{2a})_{\ovl{i}\ovl{k}} \,\mathcal{D}_{\bar{k} \ell} (u^{(m)}_{2a})_{\ell j} \right\}\varepsilon_{msn}(\tau_n)_I^{\ J}    \right].
\label{eq2.105}
\end{align}
We also obtain the following expressions 
\begin{align}
2 \sum_{a} c_{a}^{} (\bar{u}_{2a}^{(m)})_{\bar{i} \bar{k}} \, \mathcal{D}_{\bar{k} \ell} \ (u_{2a}^{(s)})_{\ell j} 
& =  \begin{pmatrix}
0 & 0 & 0 \\
-{i}{\sqrt{2}}\, \ovl{\lambda}_{2}^{(m,s)\dot{\A}} &  \ovl{\sigma}^{\mu\dot{\A}\A} A_\mu^{(m,s)} & 0 \\
- D_{2}^{(m,s)} - i\partial^\mu A_\mu^{(m,s)}  - 2i A_\mu^{(m,s)} \partial^\mu & -{i}{\sqrt{2}} \lambda_{2}^{(m,s)\A} & 0
\end{pmatrix}
\label{eq2.106}
\end{align}
\begin{align}
A_\mu^{(m,s)} &= \frac{i}{m_0^2} \sum_{a} c_{a} \left( \varphi_{2a}^{(m) *} \partial_\mu \varphi_{2a}^{(s)} - \partial_\mu \varphi_{2a}^{(m) *} \varphi_{2a}^{(s)} -i\, \ovl{\psi}_{2a}^{(m)} \ovl{\sigma}^\mu \psi_{2a}^{(s)} \right),
\label{eq2.107}
\\
\lambda_{2\A}^{(m,s)} &= -\sqrt{2} \frac{i}{m_0^2}\sum_{a} c_{a} \left(
F_{2a}^{(m)*} \psi^{(s)}_{ {2a}\A} - i \sigma^\mu_{\A\dot{\A}} \ovl{\psi}_{2a}^{(m)\dot{\A}*} \partial_\mu \varphi^{(s)}_{2a} \right),
\label{eq2.108} \\
D_2^{(m,s)} &= -\frac{1}{m_0^2}  
\sum_{a} c_{a} \left[ 2  F_{2a}^{(m)*} F_{2a}^{(s)}-2(\partial^\mu \varphi_{2a}^{(m)*} \partial_\mu \varphi_{2a}^{(s)})  \right.
\nonumber \\
&\left. \hspace{3cm} + i  \left\{ \partial_\mu \ovl{\psi}^{(m)}_{{2a}\dot{\A}} \ovl{\sigma}^{\mu \dot{\A}\A} \psi^{(s)}_{{2a}\A} -\ovl{\psi}^{(m)}_{{2a}\dot{\A}} \ovl{\sigma}^{\mu \dot{\A}\A} \partial_\mu \psi^{(s)}_{{2a}\A} \right\}  \right].
\end{align}
Substituting these into (\ref{eq2.105}), we obtain 
\begin{align}
& V_1^{[2]} = 2 \sum_a c_a(\bar{u}_{a}^{[2]})_{\bar{i} \bar{k}} \, \mathcal{D}_{\bar{k} \ell} \ (u_{a}^{[2]})_{\ell j} 
 = \begin{pmatrix}
0 & 0 & 0 \\
-{i}{\sqrt{2}}\, \ovl{\lambda}_{}^{[2]\dot{\A}} &  \ovl{\sigma}^{\mu\dot{\A}\A} A_\mu^{[2]} & 0 \\
- D_{}^{[2]} - i\partial^\mu A_\mu^{[2]}  - 2i A_\mu^{[2]} \partial^\mu & -{i}{\sqrt{2}} \lambda_{}^{[2]\A} & 0
\end{pmatrix},
\label{eq2.121}
\end{align}
where 
\begin{align}
A_\mu^{[2]} &= 2(A_\mu^{(0,0)} + \sum_m A_\mu^{(m,m)}) \frac{\tau_0}{2} +
\sum_n \left(2i (A_\mu^{(0,n)} - A_\mu^{(n,0)}) + i \sum_{m,s} \varepsilon_{nms} (A_\mu^{(m,s)}- A_\mu^{(s,m)})\right) \frac{\tau_n}{2}
\nonumber \\
&= A_\mu^{(0)}  + \sum_n \frac{\tau_n}{2}A_\mu^{(n)},
\\
 A_\mu^{(0)} &= 2 (A_\mu^{(0,0)} + \sum_m A_\mu^{(m,m)}),
\\
 A_\mu^{(n)} &= \left( 2i (A_\mu^{(0,n)} - A_\mu^{(n,0)}) + i \sum_{m,s} \varepsilon_{nms} (A_\mu^{(m,s)}- A_\mu^{(s,m)})\right).
\end{align}
In the same way, we have 
\begin{align}
\lambda_\A^{[2]} &= 2(\lambda_{2\A}^{(0,0)} + \sum_n  \lambda_{2\A}^{(n,n)}) \frac{\tau_0}{2} + \sum_n \left(2i (\lambda_{2\A}^{(0,n)} - \lambda_{2\A}^{(n,0)}) + i \sum_{m,s} \varepsilon_{nms} (\lambda_{2\A}^{(m,s)} - \lambda_{2\A}^{(s,m)})\right) \frac{\tau_n}{2}
\nonumber \\
&= \lambda_{2\A}^{(0)}\frac{\tau_0}{2} + \sum_n  \lambda_{2\A}^{(n)}\frac{\tau_n}{2},
\end{align}
\begin{align}
D^{[2]} &= 2(D_{2}^{(0,0)} + \sum_n D_{2}^{(n,n)}) \frac{\tau_0}{2} 
\nonumber \\
&\hspace{3.5cm} + \sum_n \left(2i (D_{2}^{(0,n)} - D_{2}^{(n,0)}) + i \sum_{m,s} \varepsilon_{nms} (D_{2}^{(m,s)} - D_{2}^{(s,m)}) \right) \frac{\tau_n}{2}
\nonumber \\
&= D_2^{(0)} \frac{\tau_0}{2} + \sum_n  D_2^{(n)}\frac{\tau_n}{2}.
\end{align}

As for $V_2^{[22]}$, we employ the similar operations to $V_2^{[33]}$,(\ref{V3331s})$\sim $(\ref{eq2.50c}), and obtain   
\begin{align}
& (V_2^{[22]})_{\bar{3}1} = -A_\mu^{[2]} A^{[2]\mu}, \mbox{  the other elements}= 0.
\end{align}
The $V_1^{[2]}$ and $V_2^{[22]} $ amount to the same form as (\ref{V3WZ}) replacing the upper indices $[3],[33]$ to $[2],[22]$:
\begin{align}
(V^{[2]}_{WZ})_{\bar{i}j} &= V^{[2]}_{\bar{i}j} + V'^{[22]}_{\bar{i}j}
\nonumber \\
&= 
\begin{pmatrix}
0 & 0 & 0 \\
-{i}{\sqrt{2}} \ovl{\lambda}^{[2]\dot{\A}} &  \ovl{\sigma}^{\mu\dot{\A}\A} A_\mu^{[2]} & 0 \\
- D^{[2]} - i\partial^\mu A_\mu^{[2]}  - 2i A_\mu^{[2]} \partial^\mu - A_\mu^{[2]}A^{[2]\mu} & -{i}{\sqrt{2}} \lambda^{[2]\A} & 0
\end{pmatrix}, \label{V2WZ}
\end{align}
As for the opposite chirality sector,$\ovl{V}_{WZ}^{[2]} $,  we also obtain the same form as (\ref{barVWZ3}) replacing the upper index 
$[3],[33] $ to $[2],[22]$.
\subsection{\large  U(3)$\otimes$U(2),U(3)$\otimes$U(1)}
\ \ \ 
In Eq.(\ref{VforQ}), in addition to the elements in $V^{[3]}_{WZ},\bar{V}_{WZ}^{[3]}$,$V^{[2]}_{WZ},\bar{V}_{WZ}^{[2]}$,
there exist elements which correspond to the second and forth term for ($r,r^\prime$)=($2,3$),($3,2)$. 
Under the Wess-Zumino gauge condition, we calculate these terms as follows:
\begin{align}
\frac{1}{2} [(V_2^{[23]})_{\bar{1} 1}]_{IA}^{JB} &= \sum_{a,b} c_a c_b \,[(\ovl{u}_{ab}^{[23]})_{\bar{1}\bar{k}}]_{IA}^{I'A'} \mathcal{D}_{\bar{k} \ell} \,[(u_{ab}^{[23]})_{\ell 1}]_{I'A'}^{JB} 
\nonumber \\
&= \sum_{a,b} c_a c_b [(\bar{u}_a^{[2]})_{\bar{1}\bar{1}}]_I^{I'}[(\bar{u}_b^{[3]})_{\bar{1}\bar{1}}]_A^{A'}\mathcal{D}_{\bar{1}3}\sum_k[(u_a^{[2]})_{3k}]_{I'}^J [(u_b^{[3]})_{k1}]_{A'}^B
\nonumber \\
&= \frac{1}{m_0^4} \sum c_a c_b (\varphi_a^{[2]*})_I^{I'} (\varphi_b^{[3]*})_A^{A'} \left[(F_a^{[2]})_{I'}^{J}(\varphi_b^{[3]})_{A'}^B - (\psi_a^{[2]})_{I'}^{J}(\psi_b^{[3]})_{A'}^B + (\varphi_a^{[2]})_{I'}^{J} (F_b^{[3]})_{A'}^B \right]
\nonumber \\
&= \frac{1}{m_0^4} \sum c_a c_b \left[(\varphi_a^{[2]*} F_a^{[2]})_{I}^{J}(\varphi_b^{[3]*} \varphi_b^{[3]})_{A}^B - (\varphi_a^{[2]*} \psi_a^{[2]})_{I}^{J}(\varphi_b^{[3]*} \psi_b^{[3]})_{A}^B + (\varphi_a^{[2]*}\varphi_a^{[2]})_{I}^{J}(\varphi_b^{[3]*} F_b^{[3]})_{A}^B \right]
\nonumber \\
&= 0.
\label{eq2.135d}
\end{align}
In (\ref{eq2.135d}), the contraction of indices of internal degrees of freedom corresponds to  
extracting $(3,8)$ representation in $SU(2) \otimes SU(3) \subset U(2) \otimes U(3)$
from the tensor product $(3,8) \otimes (3,8)$.
Repeating the similar calculations, we have vanishing elements but $(V_2^{[23]})_{\bar{3}1}$.
As for $(V_2^{[23]})_{\bar{3}1}$, we obtain the following expressions:
\begin{align}
\frac{1}{2} [(V_2^{[23]})_{\bar{3} 1}]_{IA}^{JB} &= \sum_{a,b} c_a c_b \,[(\ovl{u}_{ab}^{[23]})_{\bar{3}\bar{k}}]_{IA}^{I'A'} \mathcal{D}_{\bar{k} \ell} \ [(u_{ab}^{[23]})_{\ell 1}]_{I'A'}^{JB} 
\nonumber \\
&= \frac{1}{m_0^2} \sum c_a c_b \left[(F_{ab}^{[23]*})_{IA}^{I'A'} (F_{ab}^{[23]})_{I'A'}^{JB} -i (\ovl{\psi}_{ab}^{[23]})_{IA}^{I'A'} \ovl{\sigma}^\mu \partial_\mu (\psi_{ab}^{[23]})_{I'A'}^{JB} \right. 
\nonumber \\
& \left. \hspace{8cm} + (\varphi_{ab}^{[23]*})_{IA}^{I'A'} \square (\varphi_{ab}^{[23]})_{I'A'}^{JB}
\right],
\end{align}
where we have 
\begin{align}
(F_{ab}^{[23]*})_{IA}^{I'A'} &= m_0 [(\ovl{u}_{ab}^{[23]})_{\bar{3}\bar{1}}]_{IA}^{I'A'} 
= m_0 \sum_{\bar{\ell}} [(\ovl{u}_{a}^{[2]})_{\bar{3}\bar{\ell}}]_{I}^{I'} [(\ovl{u}_{b}^{[3]})_{\bar{\ell}\bar{1}}]_{A}^{A'} 
\nonumber \\
&= \frac{1}{m_0}\left[(F_a^{[2]*})_I^{I'}(\varphi_b^{[3]*})_A^{A'} - (\ovl{\psi}_{a\dot{\A}}^{[2]})_I^{I'} (\ovl{\psi}_{b}^{[3]\dot{\A}})_A^{A'} + (\varphi_a^{[2]*})_I^{I'}(F_b^{[3]*})_A^{A'} \right],
\end{align}
\begin{align}
(F_{ab}^{[23]})_{I'A'}^{JB} &= m_0 [({u}_{ab}^{[23]})_{{3}{1}}]_{I'A'}^{JB} 
= m_0 \sum_{{\ell}} [({u}_{a}^{[2]})_{{3}{\ell}}]_{I'}^{J} [({u}_{b}^{[3]})_{{\ell}{1}}]_{A'}^{B} 
\nonumber \\
&= \frac{1}{m_0}\left[(F_a^{[2]})_{I'}^{J}(\varphi_b^{[3]})_{A'}^{B} - ({\psi}_{a}^{[2]\A})_{I'}^{J} ({\psi}_{b\A}^{[3]})_{A'}^{B} + (\varphi_a^{[2]})_{I'}^{J}(F_b^{[3]})_{A'}^{B} \right],
\\
m_0^2 (F_{ab}^{[23]*})_{IA}^{I'A'} (F_{ab}^{[23]})_{I'A'}^{JB} &= (\ovl{\psi}_a^{[2]} \psi_a^{[2]})_I^J \,(\ovl{\psi}_b^{[3]} \psi_b^{[3]})_A^B + \mbox{(terms which vanish by W.Z. condition)},
\nonumber \\
&= -\frac{1}{2} (\ovl{\psi}_a^{[2]} \bar{\sigma}^\mu \psi_a^{[2]} )_I^J \, (\ovl{\psi}_b^{[3]} \bar{\sigma}_\mu \psi_b^{[3]} )_A^B,
\end{align}
\begin{align}
(\ovl{\psi}_{ab}^{[23]})_{IA}^{I'A'} &= -m_0 [(\ovl{u}_{ab}^{[23]})_{\bar{3}\bar{2}}]_{IA}^{I'A'} 
=-m_0 \sum_{\bar{\ell}} [(\ovl{u}_{a}^{[2]})_{\bar{3}\bar{\ell}}]_{I}^{I'} [(\ovl{u}_{b}^{[3]})_{\bar{\ell}\bar{2}}]_{A}^{A'} 
\nonumber \\ 
&= \frac{1}{m_0} \left[(\ovl{\psi}_{a\dot{\A}}^{[2]})_I^{I'}(\varphi_b^{[3]*})_A^{A'} + (\varphi_a^{[2]*})_I^{I'}(\ovl{\psi}_{b\dot{\A}}^{[3]})_A^{A'}  \right],
\\
({\psi}_{ab}^{[23]})_{I'A'}^{JB} &= m_0 [({u}_{ab}^{[23]})_{{2}{1}}]_{I'A'}^{JB} = m_0 \sum_{{\ell}} [({u}_{a}^{[2]})_{{2}{\ell}}]_{I'}^{J} [({u}_{b}^{[3]})_{{\ell}{1}}]_{A'}^{B}, 
\nonumber \\ 
&= \frac{1}{m_0} \left[({\psi}_{a{\A}}^{[2]})_{I'}^{J}(\varphi_b^{[3]})_{A'}^{B} + (\varphi_a^{[2]})_{I'}^{J}({\psi}_{b{\A}}^{[3]})_{A'}^{B}  \right],
\\
m_0^2 (\ovl{\psi}_{ab}^{[23]*})_{IA}^{I'A'} \ovl{\sigma}^\mu \partial_\mu (\psi_{ab}^{[23]}&)_{I'A'}^{JB} = (\ovl{\psi}_a^{[2]} \bar{\sigma}^\mu \psi_a^{[2]})_I^J \, (\varphi_b^{[3]*} \partial_\mu \varphi_b^{[3]})_A^B + (\varphi_a^{[2]*} \partial_\mu \varphi_a^{[2]})_I^{J} \, (\ovl{\psi}_b^{[3]} \bar{\sigma}^\mu\psi_b^{[3]})_A^B
\nonumber \\
&\hspace{4cm} + \mbox{(terms which vanish by W.Z. condition)},
\end{align}
\begin{align}
\nonumber \\
(\varphi_{ab}^{[23]*})_{IA}^{I'A'} &= m_0 [(\ovl{u}_{ab}^{[23]})_{\bar{3}\bar{3}}]_{IA}^{I'A'} = m_0 \sum_{\bar{\ell}} [(\ovl{u}_{a}^{[2]})_{\bar{3}\bar{\ell}}]_{I}^{I'} [(\ovl{u}_{b}^{[3]})_{\bar{\ell}\bar{3}}]_{A}^{A'} 
\nonumber \\
&= \frac{1}{m_0}(\varphi_a^{[2]*})_I^{I'} (\varphi_b^{[3]*})_A^{A'},
\\
(\varphi_{ab}^{[23]})_{I'A'}^{JB} &= m_0 [({u}_{ab}^{[23]})_{{3}{3}}]_{I'A'}^{JB} = m_0 \sum_{{\ell}} [({u}_{a}^{[2]})_{{3}{\ell}}]_{I'}^{J} [({u}_{b}^{[3]})_{{\ell}{3}}]_{A'}^{B} 
\nonumber \\
&= \frac{1}{m_0}(\varphi_a^{[2]})_{I'}^{J} (\varphi_b^{[3]})_{A'}^{B},
\\
 m_0^2 (\varphi_{ab}^{[23]*})_{IA}^{I'A'} \square (\varphi_{ab}^{[23]})_{I'A'}^{JB} &= (\varphi_a^{[2]*})_I^{I'} (\varphi_b^{[3]*})_A^{A'} \partial^\mu \partial_\mu (\varphi_a^{[2]})_{I'}^J (\varphi_b^{[3]})_{A'}^B
\nonumber \\
&= 2 \,(\varphi_a^{[2]*} \partial_\mu \varphi_a^{[2]})_I^J \, (\varphi_b^{[3]*} \partial^\mu \varphi_b^{[3]})_A^B
\nonumber \\
&\hspace{3cm} + \mbox{(terms which vanish by W.Z. condition)}.
\end{align}
In result, we obtain
\begin{align}
[(V_2^{[23]})_{\bar{3} 1}]_{IA}^{JB} &= \frac{2}{m_0^4} \sum_{a,b} c_a c_b \,\left[ -\frac{1}{2} (\ovl{\psi}_a^{[2]} \bar{\sigma}_\mu \psi_a^{[2]} )_I^J \, (\ovl{\psi}_b^{[3]} \bar{\sigma}^\mu \psi_b^{[3]} )_A^B -i \,(\ovl{\psi}_a^{[2]} \bar{\sigma}_\mu \psi_a^{[2]})_I^J \, (\varphi_b^{[3]*} \partial^\mu \varphi_b^{[3]})_A^B 
\right.
\nonumber \\
& \left. \hspace{2.5cm} -i \, (\varphi_a^{[2]*} \partial_\mu \varphi_a^{[2]})_I^{J} \, (\ovl{\psi}_b^{[3]} \bar{\sigma}^\mu\psi_b^{[3]})_A^B + 2 \,(\varphi_a^{[2]*} \partial_\mu \varphi_a^{[2]})_I^J \, (\varphi_b^{[3]*} \partial^\mu \varphi_b^{[3]})_A^B 
\right]
\nonumber \\
&= - (A_\mu^{[2]})_I^J (A^{[3] \mu})_A^{B}
\nonumber \\
&= [(V_2^{[32]})_{\bar{3} 1}]_{IA}^{JB}, \label{V32}
\end{align}
where the last equation is due to the commutativity of $(A_\mu^{[2]})_I^J $ and $(A^{[3] \mu})_A^{B}$. 

Continuing the similar calculations for the opposite chirality sector, we obtain $\bar{V}_2^{[23]}$ expressed by
\begin{align}
& [(\ovl{V}_2^{[23]})_{\bar{3} 1}]_{IA}^{JB} = [(\ovl{V}_2^{[32]})_{\bar{3} 1}]_{IA}^{JB}= - (A_\mu^{[2]})_I^J (A^{[3] \mu})_A^{B}, 
\label{barV32}\mbox{  the other elements} =0. 
\end{align}

Adding $\mathcal{D}$,$\ovl{\mathcal{D}}$ and fluctuations due to the algebra which act on $Q$ field, we obtain the 
fluctuated Dirac operator $\widetilde{\mathcal{D}}_Q$, $\widetilde{\ovl{\mathcal{D}}}_Q$ in (\ref{tDQ}) and (\ref{tbarDQ}). 
The derived vector supermultiplet become to be adjoint representation of $U(3)\times U(2)$.

To the fluctuation for $\mathcal{D}_M$ in the $u_R,d_R$ sector due to the elements expressed by Eq.(\ref{UuRdR}), 
similar operations carry the 
same result as (\ref{V32}),(\ref{barV32}) but replacing the index [32] to [31] so that we obtain (\ref{DuRdR}) in which 
the vector supermultiplet($A_\mu^{(s)},\lambda_\A^{(s)},D^{(s)}$),$s=u_R(d_R)$ becomes to be adjoint representation of 
$U(3)\times U(1)$.


\begin{thebibliography}{99}
\bibitem{rf:Grcoupled}
A. Connes,Comm. Math. Phys. {\bf 182},155(1996), hep-th/9603053.
\bibitem{rf:NCGneutrino} 
A.H.Chamseddine, A.Connes, and M.Marcolli,hep-th/0610241,2006.
\bibitem{rf:JHigh}
A.Connes, J.High Energy Phys.{\bf 0611},081(2006).
\bibitem{connes0} A.Connes, Noncommutative Geometry, Quantum Fields and Motives, 2008.
\bibitem{connes7}
A.Connes, Comm. Math. Phys. {\bf 182},155(1996), hep-th/9603053.
\bibitem{connes10}
A. H. Chamseddine and A. Connes, J. Geom. Phys, {\bf 57},1(2006), hep-th/0605011.
\bibitem{connes8}
A.H.Chamseddine, A.Connes, Comm. Math. Phys. {\bf 186},731,(1997),hep-th/9606001.
\bibitem{paper0}
S. Ishihara, H.Kataoka, A.Matsukawa, Hikaru Sato and M.Shimojo, arXiv:1311.4666(hep-th).
\bibitem{paper1}
S. Ishihara, H.Kataoka, A.Matsukawa, Hikaru Sato and M.Shimojo, arXiv:1311.4671(hep-th).
\bibitem{martin}
S.P.Martin, "A supersymmetry primer",(2008),hep-ph/9709355.
\end{thebibliography}

\end{document}